# On beam characterization of ground-based CMB radio telescopes using UAV-mounted sources: application to the QUIJOTE TFGI and plans for LSPE-Strip

**Fabio Paonessa,**[a,1] **Lorenzo Ciorba,**[a,z,2] **Giuseppe Addamo,**[a] **Paz Alonso-Arias,**[b] **Barbara Caccianiga,**[c,y] **Marco Bersanelli,**[c] **Francesco Cuttaia,**[d] **Cristian Franceschet,**[c] **Ricardo Tanausú Génova Santos,**[b] **Massimo Gervasi,**[e,x,w] **Roger Hoyland,**[b] **Mike Jones,**[f] **Carlos Hugo López-Caraballo,**[b] **Mauro Lumia,**[a] **Michele Maris,**[g,*,**] **Aniello Mennella,**[c] **Gianluca Morgante,**[d] **Oscar Antonio Peverini,**[a] **Sabrina Realini,**[c,y] **José Alberto Rubiño-Martín,**[b] **Stefano Sartor,**[g] **Angela Taylor,**[f] **Fabrizio Villa,**[d] **Mario Zannoni**[e,w] **and Giuseppe Virone**[a]

[a] *Consiglio Nazionale delle Ricerche, Istituto di Elettronica e di Ingegneria dell'Informazione e delle Telecomunicazioni (CNR-IEIIT),*
*Corso Duca degli Abruzzi, 24, 10129 Turin, Italy*

[b] *Instituto de Astrofísica de Canarias (IAC),*
*C/ Vía Láctea, s/n E-38205 La Laguna, Tenerife, Spain*

[c] *Università degli Studi di Milano, Dipartimento di Fisica,*
*Via Celoria, 16, 20133 Milan, Italy*

[d] *Istituto Nazionale di Astrofisica (INAF), Osservatorio di Astrofisica e Scienza dello Spazio (OAS),*
*Via Gobetti, 93/3, 40129 Bologna, Italy*

[e] *Dipartimento di Fisica, Università di Milano Bicocca,*
*Piazza della Scienza, 3, 20126 Milan, Italy*

[f] *University of Oxford,*
*Denys Wilkinson Building, Keble Road, Oxford OX1 3RH, UK*

[g] *Istituto Nazionale di Astrofisica (INAF), Osservatorio Astronomico di Trieste (OATs),*
*Via Giambattista Tiepolo, 11, 34143 Trieste, Italy*

[w] *Istituto Nazionale di Fisica Nucleare (INFN), Sezione di Milano Bicocca,*
*Piazza della Scienza, 3, 20126 Milan, Italy*

[x] *Conseil Européen pour la Recherche Nucléaire (CERN),*
*Esplanade des Particules 1, P.O. Box 1211 Geneva 23, Geneva, Switzerland*

[y] *Istituto Nazionale di Fisica Nucleare (INFN), Sezione di Milano,*
*Via Celoria, 16, 20133 Milan, Italy*

[1] Corresponding author.
[2] The author was with CNR-IEIIT when this research was carried out.

[z] *University of Bern, Institute of Applied Physics,*
*Sidlerstrasse 5, 3084 Bern, Switzerland*

[*] *Institute for Fundamental Physics of the Universe (IFPU),*
*Via Beirut, 2, 34151 Trieste, Italy*

[**] *Fondazione ICSC, Centro Nazionale di Ricerca in HPC, Big Data and Quantum Computing,*
*Via Magnanelli, 2, 40033 Casalecchio di Reno (BO), Italy*

*E-mail:* fabio.paonessa@cnr.it

Abstract: The Large Scale Polarization Explorer (LSPE) project, funded by the Italian Space Agency (ASI), includes the development of LSPE-Strip, a ground-based radio telescope for observing Cosmic Microwave Background (CMB) anisotropies. LSPE-Strip, nearing its construction phase, will operate from the Teide Observatory in Tenerife, employing 49 coherent polarimeters at 43 GHz to deliver critical data on CMB anisotropies and 6 channels at 95 GHz as atmospheric monitor. On-site characterization of such advanced instruments is crucial to detect possible systematic effects, such as gain fluctuations, beam distortions, and pointing errors, that can compromise performance by introducing spurious polarizations or radiation collection from unintended directions. To address these challenges, a drone-mounted Q-band test source for on-site characterization of LSPE-Strip's polarimeter array was developed. Modern Unmanned Aerial Vehicles (UAVs) offer a flexible approach for antenna pattern measurements, yet their use in high-frequency radio astronomy is not consolidated practice. In October 2022, a UAV-based measurement campaign was conducted with the TFGI instrument on the second QUIJOTE telescope in Tenerife, in collaboration with the Instituto de Astrofísica de Canarias. This pioneering effort aimed to validate UAV-based beam characterization methods and assess QUIJOTE's performance under operational conditions. Preliminary results demonstrated high measurement accuracy, leveraging QUIJOTE's dual-receiver configuration for beam validation. These findings provide valuable insights for optimizing UAV systems in preparation for LSPE-Strip's future characterization.

## Contents

## 1 Introduction

LSPE-Strip [1] will be a ground-based radio telescope for Cosmic Microwave Background (CMB) observations, currently approaching the installation phase within the initiative named Large Scale Polarization Explorer (LSPE) funded by the Italian Space Agency (ASI). To provide key information on the CMB anisotropies and Galactic foreground emission in polarization, LSPE-Strip relies on a cluster of 49 coherent polarimeters at 43 GHz [2, 3] that will observe the sky from the Teide Observatory, Tenerife.
The on-site characterization represents a crucial phase for such complex instruments. A number of systematic effects can occur, to the detriment of the system performance. These systematic effects include receivers' gain fluctuations, beam pattern distortions, unexpected level of secondary lobes, pointing errors and many others. Ultimately, they can introduce spurious polarizations, or cause radiation being collected from uncontrolled directions, and leakages between the Stokes parameters.

Unmanned Aerial Vehicles (UAVs) have revolutionized in-situ antenna measurements by providing versatile platforms for characterizing radiation patterns across several frequency bands. While the ability of UAV-based systems to perform accurate and reliable measurements has been widely documented for low-frequency radio astronomy [4–7], the applicability of UAVs for antenna characterization in higher frequency bands, including those necessary for advanced microwave cosmological research, still presents challenges.

In the framework of the LSPE-Strip development, the IEIIT institute of the Italian National Research Council (CNR-IEIIT) developed an innovative UAV-mounted test source operating in the Q-band [8]. This technology will be used to validate the ground-based cluster of Q-band coherent polarimeters that compose LSPE-Strip, and to calibrate its Star Tracker [9]. Using a linearly polarized artificial signal source such as a UAV provides increased Signal-to-Noise Ratio (SNR) compared to celestial sources. Unlike celestial sources, which are limited in availability, position, and polarization control, a drone-based system offers high flexibility, repeatability, and tailored signal injection, making it particularly well-suited for assessing the full polarimetric response of the instrument. This approach enables the detection of systematic effects such as beam distortions and pointing inaccuracies [10].
In October 2022, a measurement campaign was conducted on a similar instrument, the TFGI instrument installed on the second Q-U-I JOint TEnerife (QUIJOTE) telescope, located at the Teide Observatory in Tenerife [11]. The campaign on QUIJOTE-TFGI, carried out in collaboration with the Instituto de Astrofísica de Canarias (IAC), had a dual relevant purpose: to provide valuable results concerning the telescope's performance verification under actual operating conditions——a task never previously undertaken with a drone——and to gather insights into the UAV system's performance that will support its optimization in view of the future LSPE-Strip characterization.

This paper is organized as follows: Section 2 outlines the main features of LSPE-Strip and QUIJOTE, while Section 3 describes the key characteristics of the UAV subsystem. Section 4 provides an overview of the measurement campaign, including adopted strategies and data processing workflow. The most significant results are presented and discussed in Sections 5 and 6. Finally, Section 7 offers considerations on the campaign's main outcomes and projections for the future characterization of LSPE-Strip, including anticipated changes.

## 2 LSPE-Strip and QUIJOTE-TFGI: similarities and differences

Both LSPE-Strip [1] and QUIJOTE [12, 13] are ground-based CMB telescopes with an elevation-over-azimuth pointing system. LSPE-Strip will be installed in proximity of the QUIJOTE telescopes, therefore, all the considerations on measurement logistics and scan strategy discussed in section 4 are applicable to both instruments. The main mirror of a QUIJOTE telescope has a projected aperture of 2.25 m [12] with a Full Half-Power Beam Width (FHPBW) of 22.5 and 18.4 arcmin at 30 and 40 GHz, respectively. LSPE-Strip has instead a main mirror of 1.5 m providing a FHPBW of 19.8 arcmin at 43 GHz [1]. These narrow beams represent a significant challenge for UAV-based antenna measurement systems. The FHPBW values are significantly smaller than those previously reported in the literature [14, 15], except for near-field approaches on large reflector antennas [16]. These works, however, rely on tethered drones and advanced laser-based tracking systems with constraints of limited tracking volume and high susceptibility to challenging environmental conditions (e.g., intense solar radiation), making them unsuitable for the characterization of instruments such as LSPE and QUIJOTE.
In the configuration available in 2022, the second QUIJOTE telescope had a focal plane with 7 pixels grouped in two receivers working at 30 GHz (Thirty Gigahertz Instrument, TGI) and 40 GHz (Forty Gigahertz Instrument, TGI). The combined set of receivers is named TFGI instrument. This feature already allowed to develop a multi-beam measurement technique (see section 6) in view of the wider focal plane of LSPE-Strip (49 polarimeters in Q-band and 6 in W-band). Both instruments feature correlation polarimeter receivers [17, 18] and operate in dual-circular polarization at antenna level. Each correlation polarimeter is mainly composed of a horn [3], a waveguide polarizer [19], an orthomode transducer [20] and a correlation unit. The LSPE-Strip polarimeters feature a highly-integrated correlation unit module [21] that is directly connected behind each orthomode transducer inside the cryostat. On the contrary, the QUIJOTE polarimeters are split into a cold front end (from horn to first Low-Noise-Amplifier) and a warm back end [13].

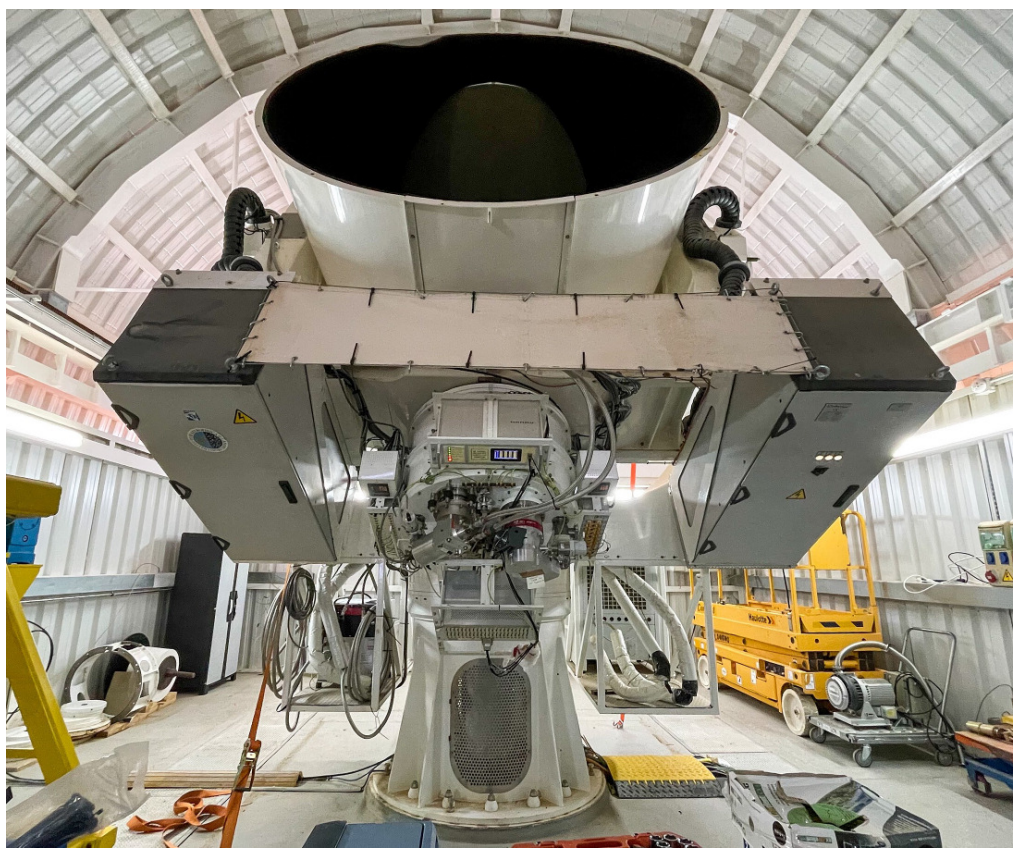

(a)

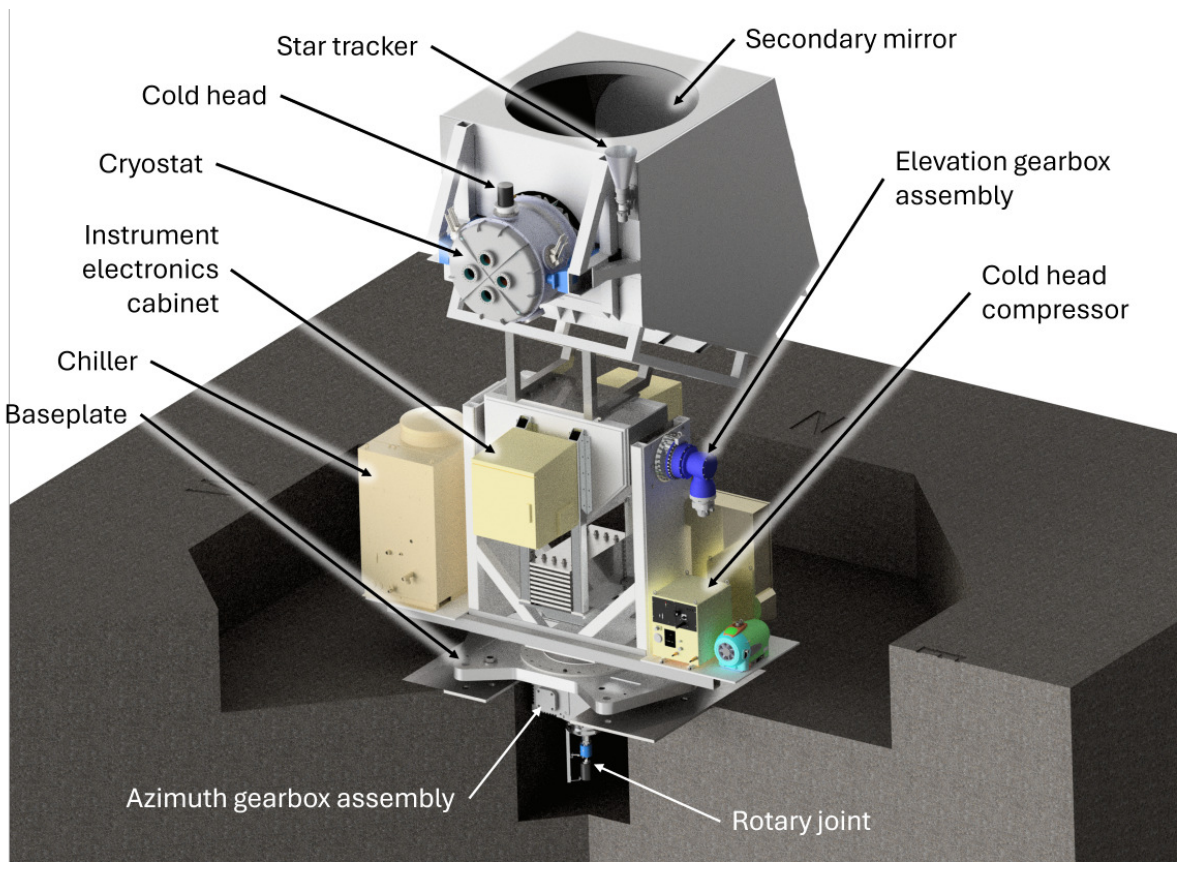

(b)

**Figure 1**. (a) The second QUIJOTE telescope (QT2) and TFGI instrument. Photograph by Mike Peel (`https://www.mikepeel.net`, CC BY-SA 4.0, via Wikimedia Commons. (b) Rendering of LSPE-Strip.

Throughout the QUIJOTE test campaign we executed two different type of measurements. The architecture of QUIJOTE (see Figure 2 of [13]) allows the front-end amplifier to be disconnected from the downstream signal chain. This feature enables to measure the received signal at both front-end output, by using a spectrum analyzer, and at the polarimeter (digitized) outputs. While the former is more similar to a standard antenna measurement, the latter represents is closer to an end-to-end characterization of the instrument. These two measurement configurations are complementary and lead to different considerations:

- a narrow-band receiver at the front-end level (results in Section 5) enhances the dynamic range, assuming the UAV transmits a narrow-band signal, and facilitates verification at the antenna level by isolating potential issues that may arise at subsequent stages;
- on the other hand, at polarimeter level, all the signals are measured simultaneously for all pixels in the focal plane, thus allowing the verification of beam equalization and relative pointing (results in Section 6).

## 3 UAV Subsystem

The UAV subsystem comprises the unmanned flying platform that transports the RF payload onboard, along with the ground control station necessary for real-time monitoring and control of the UAV mission. The main UAV characteristics from the point of view of navigation performance are described in Section 3.1, while the characteristics of the RF payload are explained in Section 3.2.

### 3.1 UAV Flight Characteristics

The adopted UAV is a custom version of the Italdron 4HSE [22] quadcopter model, which uses the open source Pixhawk 2.1 flight controller [23] running the ArduCopter firmware [24]. With regard to antenna measurements, the main features of the UAV are the performance of the positioning sensor and the attitude sensor. In our model, UAV positioning relies on a Real-Time Kinematic (RTK) receiver, specifically a u-blox ZED-F9P module, which provides position data with a nominal accuracy of 1 cm for both the horizontal and the vertical component [25]. The UAV attitude (i.e., the state of the three angles that describe the spatial orientation of the aircraft, namely yaw, pitch, and roll) is instead measured through the flight controller's internal Inertial Measurement Unit (IMU) in conjunction with an external RM3100 magnetometer [26]. Concerning the attitude accuracy previous research works revealed a value of a few degrees [27].

The flight execution is entirely autonomous in all its phases, from takeoff to landing. The flight plan is defined through a set of instructions describing the 3D GPS position of waypoints at which the UAV shall navigate or loiter, as well as conditional commands such as the particular heading that the UAV is required to maintain. The heading expresses the direction in which the front of the aircraft is pointed, typically measured as an angle from the true north. Generating flight plans requires *a priori* knowledge of the telescope rotation center, details of this aspect are reported in Appendix A.

Both position and attitude data, collectively referred to as *pose*, are recorded along with their corresponding timestamps at a sampling frequency of 5 Hz. The presence of timestamps is crucial, because post-flight data processing requires alignment of pose data with the RF data. The latter, it is important to note, must also be associated with timestamps.

The high accuracy of the tracking system enables effective post-flight data processing. It is important to acknowledge, however, that the UAV flight performance is subject to external factors (such as wind, temperature fluctuations, and pressure variations) and is generally less reliable than the navigational sensors employed in the flight control loop. Nonetheless, employing an RTK positioning sensor certainly leads to improved flight precision.

### 3.2 Radio Frequency Generator

For the characterization of LSPE-Strip, a specific RF payload for unmanned aircraft was developed. The design and validation phases addressed critical challenges such as frequency instability caused by thermal variations and motor vibrations, these aspects are thoroughly reported in [8], however, the main features are hereby described for better clarity.

The payload consists of a continuous-wave frequency generator with monitoring of the emitted power. As shown in Figure 2a, the generator is composed of a Valon Technology 5015 frequency synthesizer [28] that produces a clipped-sine waveform in the frequency range 8.25–12.5 GHz, followed by a Sage Millimeter (now Eravant) SWD-series active frequency multiplier [29] that converts the signal to Q-band (33–50 GHz) and feeds it into a WR-22 rectangular waveguide. To cope with possible fluctuations of the output power, which is highly dependent on the multiplier temperature, a diode-based power detector is also present onboard. System tests revealed that, with proper thermal dissipation, the power variation is negligible (<0.05 dB); however, under poor thermal dissipation conditions, the power generated by the source undergoes a significant transient, with a variation of up to 2 dB. The high-power signal is routed to the detector's input through a waveguide coupler. The detector voltage is read and converted to a digital format by the flight controller, specifically via an auxiliary Analog to Digital Converter (ADC) input. The time-stamped voltage readings are stored into the internal log-file along with both pose data to allow data alignment and post-flight elaboration. Power readings are used to compensate for distortions that can occur during the flight. The generated signal is radiated through a rectangular horn antenna having a maximum gain of 15 dBi. Figure 2b shows that the onboard horn points at 60 degrees from nadir toward the forward direction of the UAV. Within the vertical symmetry plane of the UAV, the polarization is linear and lying in the same plane.

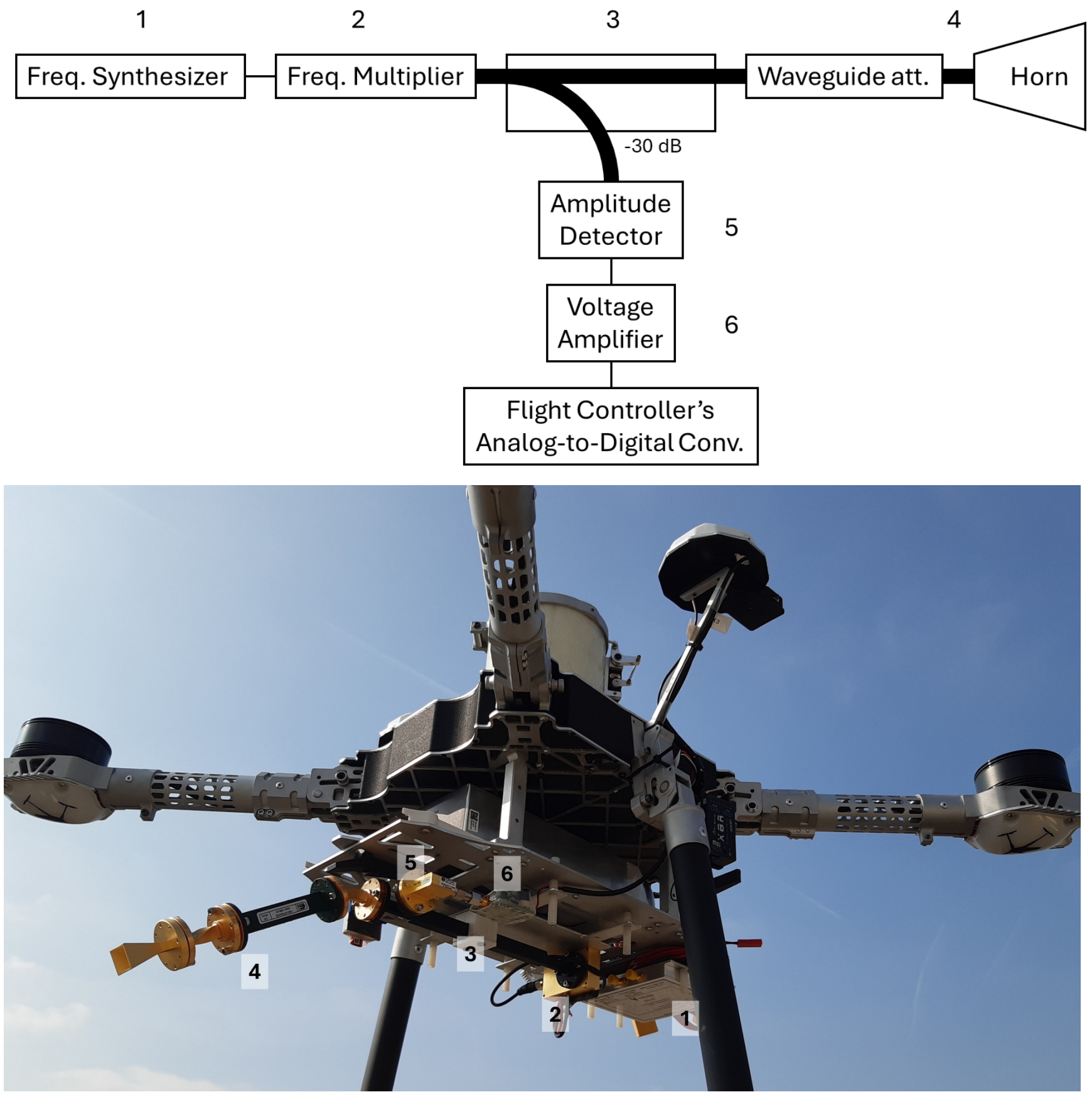

(a)

(b)

**Figure 2**. Scheme (a) and photograph (b) of the RF payload. Numbered items: (1) frequency synthesizer; (2) ×4 frequency multiplier; (3) directional coupler; (4) bend, attenuator, twist, and horn; (5) amplitude detector; (6) precision voltage amplifier on ad-hoc printed circuit board. Waveguide connections drawn with thick lines, wired connections with thin lines.

## 4 Measurement Requirements, Strategies and Data Processing

As stated in the introduction, the primary goal of the measurement campaign was the verification of the telescope beams using a drone-based approach. The scanning strategy had to account for various factors, including the drone's flight performance and the time required to complete the scan. The aim is to sample the beam over a 2D surface, from which cuts along the principal planes (E-plane and H-plane of the radiation pattern) could be extracted. These cuts provide a more straightforward comparison between measured and simulated data.

Figure 3 shows a map view of the Teide Observatory site in Tenerife, where QUIJOTE is located and that will also host LSPE-Strip. The figure highlights the chosen takeoff and landing points for the UAV (home location, top-left green marker) and the designated hovering point (bottom-left green marker) where the UAV was programmed to maintain a stationary position at an altitude of 113 meters above home location. The so-called Region Of Interest (ROI), marked by the red marker in Figure 3 and corresponding to the telescope's rotation center, was set in order for the UAV to keep its heading angle pointed toward the telescope. The coordinates of the defined hovering point correspond to 260° (azimuth) and 34° (elevation) of the telescope, at a distance of 206 meters from the telescope to the UAV.

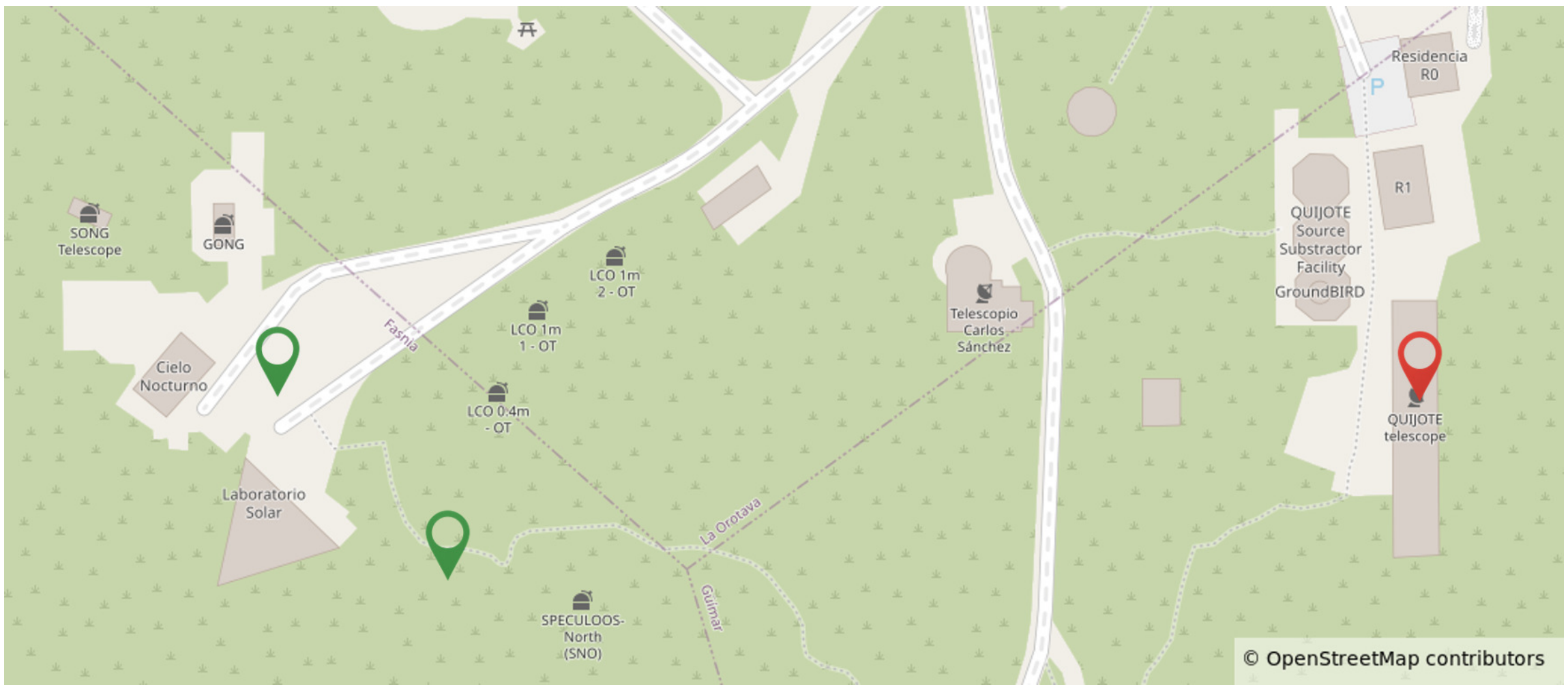

**Figure 3**. Map of the measurement site. Reference points: home location (left green marker); designated hovering point (bottom-left green marker); telescope's azimuth rotation center (red marker). Map data from OpenStreetMap.

The chosen measurement strategy involved raster scans during which the UAV hovered at the predefined waypoint while the telescope performed small alternating azimuth rotations mixed with elevation steps, as illustrated in the diagram in Figure 5. The UAV is programmed to remain constantly oriented toward the telescope, even in the presence of drone fluctuations. Compared to commonly employed strategies where the AUT is fixed and the UAV moves in the sky [30], the chosen approach ensures more controlled UAV position and orientation throughout the entire measurement process. A misalignment of the onboard horn, if limited (e.g., due to mounting tolerances), results in a negligible reduction in antenna gain in the direction of the telescope. This constitutes a constant factor that does not affect the acquired measurement.

Owing to the high directivity of the onboard horn antenna and the absence of scattering sources within its field of view, the far-field distance of the UAV-mounted test-source is on the order of 100 mm, i.e., well below the distance between the telescope and the flying drone. As far as the telescope is concerned, the far-field distance of QUIJOTE is always larger than 1000 m for the 30 and 40 GHz (TFGI) receivers. This means that the RF data are acquired in the near-field of the telescope. In its most common use, the term near-field measurements typically refers to measurements carried out in anechoic chambers with vectorial RF signal acquisition systems. Data are collected in the near field and subsequently transformed—leveraging phase information—into the far-field radiation pattern. This approach has also been applied to the characterization of low-frequency radio telescopes (outdoor), with suitable strategies adopted to enable phase measurement in systems lacking coherence between transmitter and receiver [6, 31, 32]. In the present campaign, although the measurements are still performed in the near field, the objective is different: instead of transforming the data to obtain a far-field pattern, the acquired signals are directly compared with a simulated field distribution on the same scanning surface, requiring amplitude-only measurements. This method still enables validation of the telescope model without requiring phase information. Moreover, it is the only viable approach here, as phase cannot be reliably measured not even through a reference antenna as done in [6, 32], and the far-field distance would exceed regulatory and technical constraints. The telescope model is implemented in GRASP® (see sketch in Figure 6 and additional details in [2]) and includes the nominal reflectors and the models of the corrugated feed-horns, as simulated with the CST Studio Suite® and SRSR-D 4.5®.

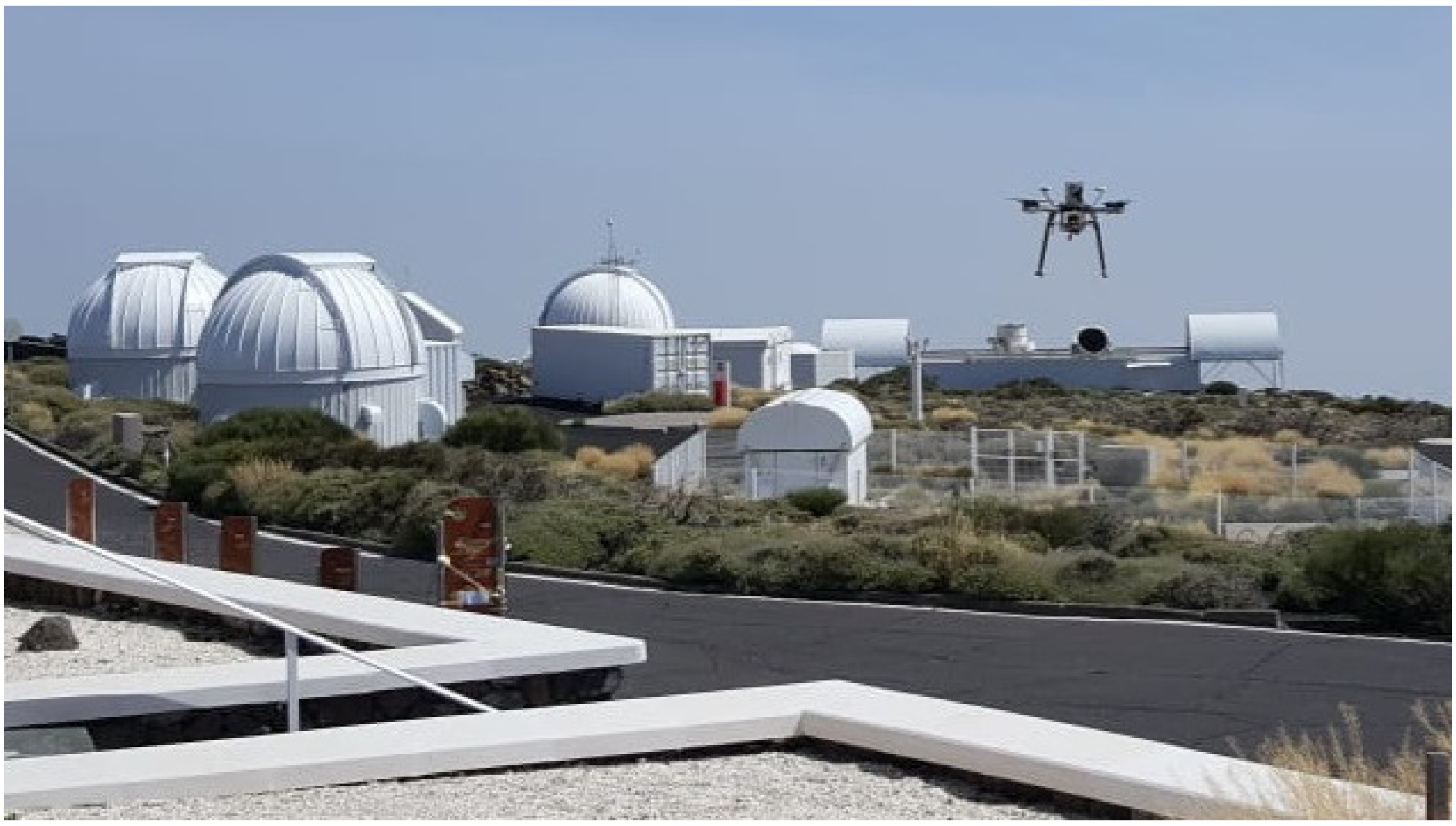

**Figure 4**. Photograph of the telescope and the UAV during the landing phase, taken near the home location. The black baffle of the second QUIJOTE telescope (the one on the right) is clearly visible beneath the UAV. The first QUIJOTE telescope is parked, pointing to zenith in the image.

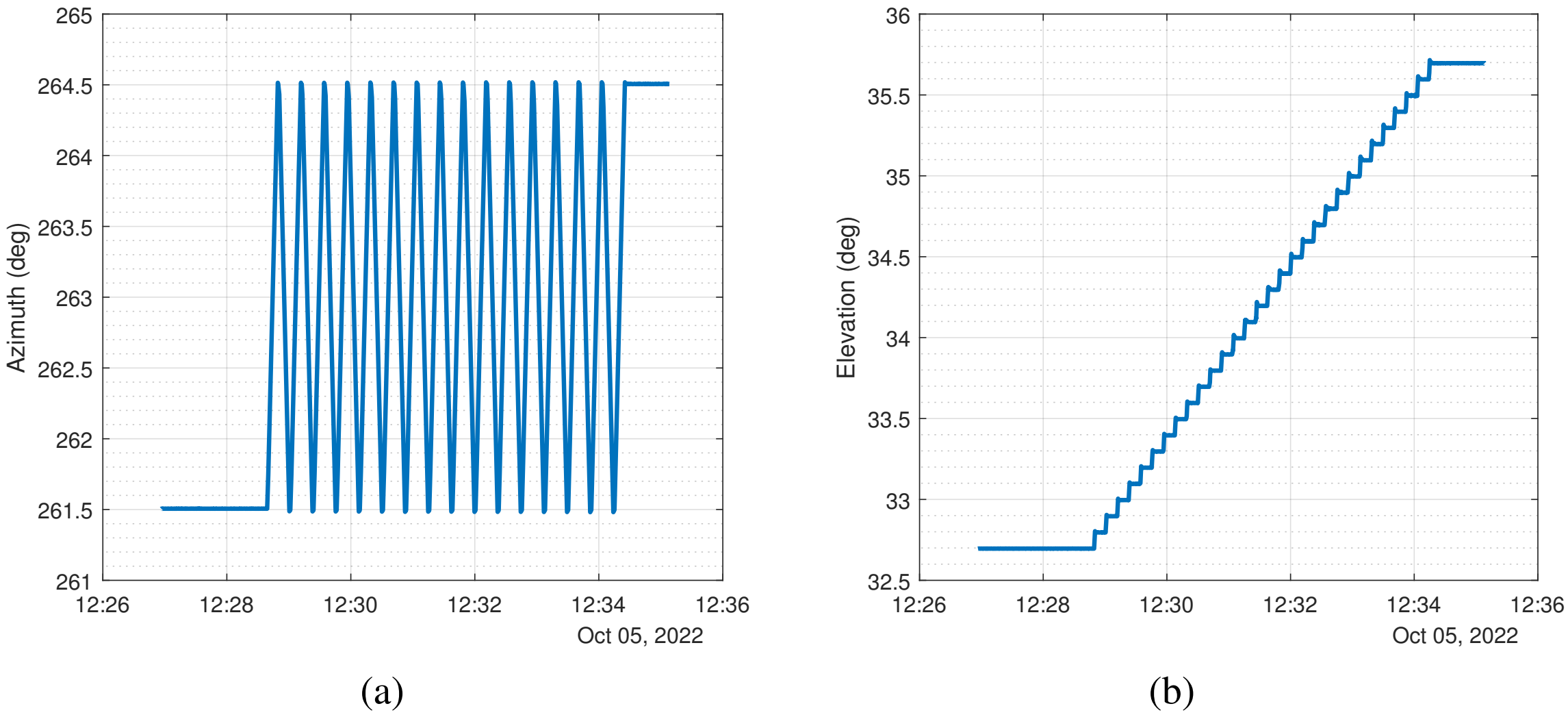

**Figure 5**. (a) Azimuth over time and (b) elevation over time of a raster scan performed with the telescope, consisting of 31 forward/backward azimuth rotations and 31 elevation steps that cover a total angular range of 3° for both angles.

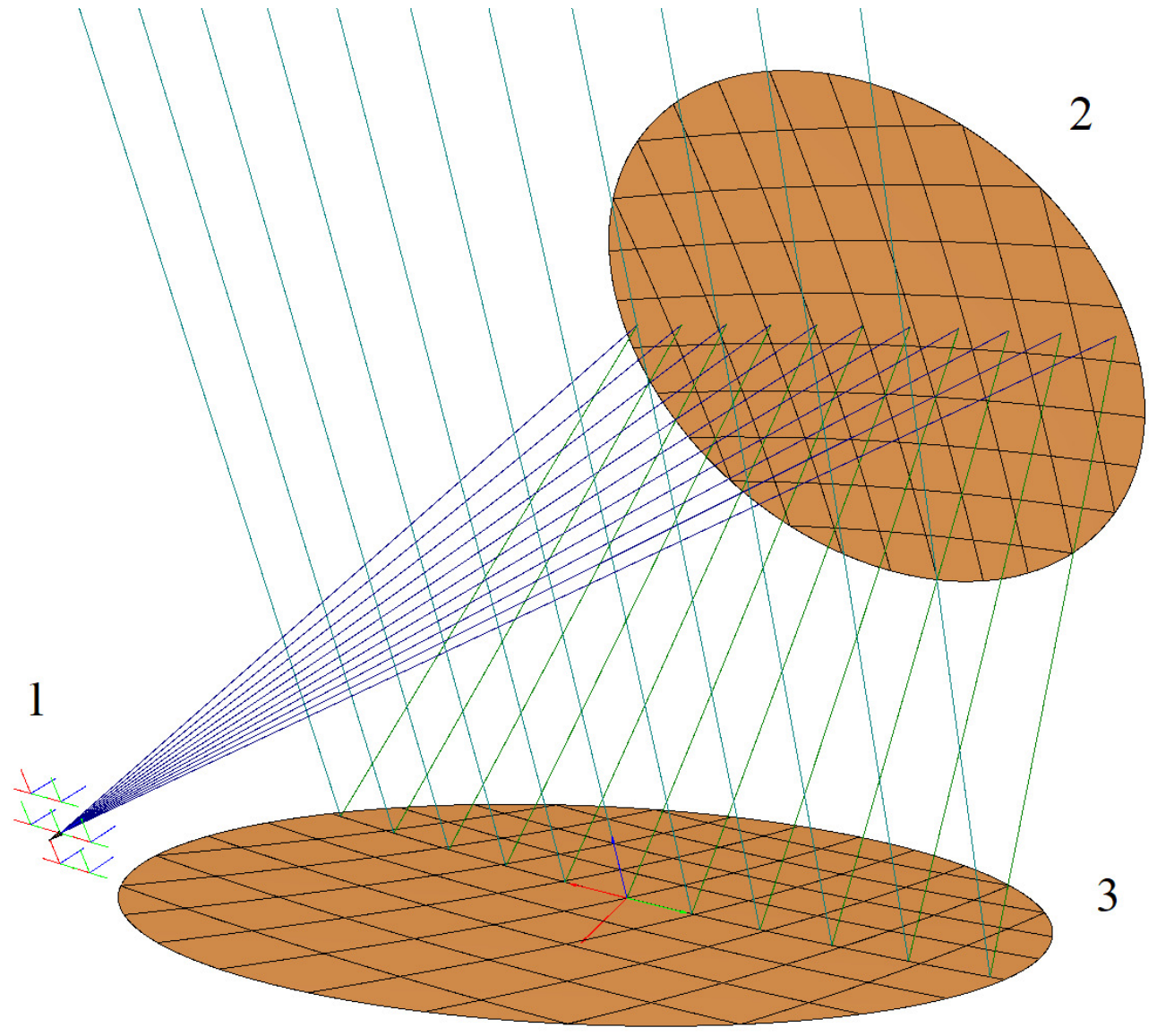

**Figure 6**. Picture of the GRASP® model used for the analysis. (1) cluster of 7 pixels of the focal plane in 2022; (2) sub-reflector; (3) main dish.

As far as measurement accuracy is concerned, no specific requirement has been set in this campaign regarding calibration accuracy on beam or other instrument parameters. The campaign on QUIJOTE serves as a first exploratory study of the potentiality of calibration campaigns based on UAV-mounted sources, which may be the only way to calibrate CMB experiments to the levels required to set sensible constraints on the B-mode signal. Recent studies performed in the context of the LiteBIRD collaboration [33] indicate that to reach the planned sensitivity of this mission, knowledge of the beam down to even the –80 dB level may be required for some angular distances.

Similarly, the polarization direction must be calibrated also to a very high precision, of the order of one arcmin [34]. It should be noted that, while this final requirement is orders of magnitude beyond the capabilities of the drone used in this campaign, advanced sensors yet not fully integrated into flight controllers are already approaching the required performance [35].

Given the aforementioned flight strategy, the RTK provides an angular accuracy of $\epsilon \approx \arctan(0.02/206) \approx 0.006°$ corresponding to 21.6 arcsec., i.e. approximately 1/60 the beamwidth, which is adequate. For QUIJOTE, the tracking error is on the order of 50 arcsec for azimuth rotation speed of 1 deg/s and 90 arcsec for 3 deg/s. For LSPE-Strip, estimation is 20–100 arcsec in azimuth and 300–1800 arcsec in elevation. A more accurate characterization of the telescope pointing parameters can be achieved by increasing the distance between the telescope and the drone in flight. This was not done during this campaign, as it would require special permits to exceed the 120-meter altitude limit above ground level.

It is worth mentioning that datasets involved in the measurements are generated from up to three independent sources, namely the telescope, the UAV, and the spectrum analyzer. Data processing therefore copes with simultaneously moving objects and uncorrelated timestamps. The post-flight processing is therefore articulated in the following phases:

1. Data interpolation: UAV's pose data are interpolated into the RF time vector, the latter is generated by either the telescope (measurements at polarimeter outputs) or the spectrum analyzer (measurements at front end level). In the second case, also the pointing data of the telescope are interpolated into the RF time vector.

2. Position and pointing data merging: telescope's encoder readings and UAV position data—both expressed in azimuth and elevation (sky coordinates)—are merged together to compute the UV coordinates of the test-source in the feed reference frame using the telescope pointing model [11]. The computed UV set represents the actual sampling points in the near-field beam. See Appendices A and B for details on telescope's rotation center determination and reference system transformation, respectively.

3. Simulated near-field beams are computed in the aforementioned actual sampling points.

Throughout the measurement campaign, a total of 10 flights were accomplished. The effective duration, defined as the hovering time at the predefined point, ranged from 6 minutes to 6 minutes and 30 seconds. The most significant results are presented and discussed in Sections 5 (at front-end) and Section 6 (polarimeter measurements).

## 5 Results at Front-End Output

The measurements at the output of the front-end were acquired using an Anritsu MS2760A spectrum analyzer connected via coaxial cable to Feed #1. This portable analyzer was housed within the telescope's rack alongside a laptop and a u-blox LEA-6T [36] timing GPS receiver. The latter provided the Pulse Per Second (PPS) signal used to trigger the spectrum analyzer, and streamed the timestamps to the laptop. Through a custom user interface developed in MATLAB®, the laptop downloaded the data from the analyzer and assigned the correct timestamps. Measurements in this configuration were conducted at a frequency of 33 GHz, corresponding to the center frequency configured for the analyzer. The resolution bandwidth was set to 1 MHz. The analyzer operated in zero-span mode, meaning the acquired trace represented the Front End Module (FEM) output as a function of time. Sweeps were performed every 500 milliseconds, with the analyzer triggered every second, resulting in 50% acquisition duty cycle. Each trace contained 501 samples, corresponding to a sampling frequency of 1 kHz.

Figure 7 presents the elevation versus azimuth for a specific flight, shown at two different zoom levels for both the telescope (black trace) and the UAV (blue trace). The azimuth rotation speed was 0.3 degrees per second. These plots reveal that the UAV is not perfectly stationary during the scan, exhibiting some movement in the sky mainly due to wind gusts. Notably, the positional variation on the horizontal plane is considerably smaller than the UAV's altitude variation. Specifically, the UAV moved approximately 2 meters vertically and about 1 meter horizontally during the flight. The higher vertical variation arises because the UAV uses RTK positioning to control its horizontal position, while the flight altitude is maintained through the barometric altimeter. This sensor is quite subject to fluctuations and resets its pressure reference at startup. As a result, changes in atmospheric pressure between flights lead to different altitude estimates, despite identical flight commands.
As outlined in Section 4, the UV coordinates are derived from the UAV's pose data and telescope pointing information. Figure 8 shows a color-coded line representing the power received by the spectrum analyzer, measuring the front end output for Feed #1 at 33 GHz. The line is discontinuous because the spectrum analyzer acquired data 50% of the time. Additionally, the epochs are neither parallel nor equally spaced due to the UAV's simultaneous movement during the telescope scan. The pink highlight indicates the set of ten chronologically consecutive epochs used to process the data shown in Figure 9. This figure compares the measured power with the simulated received power for the same UV points that were actually sampled. They are very close to the H-plane of the beam pattern. The simulation is normalized to 0 dB. The shape of the measured curve was consistent with 1 dB compression near the peak. Therefore, the data were first normalized to –1 dB, and the samples affected by compression were discarded to ensure consistency with the simulated beam profile. This allows for a quick check at the antenna level, deferring more detailed considerations to the acquisitions at polarimeter outputs (which also provided in an uninterrupted acquisition). In this configuration, a dynamic range of about 30 dB was achieved.

In order to increase both dynamic range and beam sampling, and to expand the scanning area, several flights were executed with different power levels of radiated signal, then the multiple acquisitions have been merged into a single dataset. Power levels were adjusted by adding waveguide attenuators, which had been characterized in the laboratory at the operating frequency. This allowed for an accurate merging of the datasets. Moreover, to prevent compression at higher power levels and contamination from background signals at lower levels, each set of measurements was filtered to retain only the samples within the linearity range, while discarding those affected by compression or background noise. As for system fluctuations (e.g., output power instability of the transmitter), the relatively cool ambient temperature during the measurement campaign, combined with the efficient heat dissipation of the onboard electronics, ensured that the transmitted power — monitored via the power detector onboard the drone — remained stable within 0.05 dB, which is negligible compared to all other sources of uncertainty. The overall trajectories are illustrated in Figure 10a, black line representing the telescope scan, and blue line representing the UAV trace. It is worth noticing that the collection of UAV traces, compared to the one in Figure 7a appears further elongated in elevation. This arises because of the long-term drift of the barometer response. In other words, the flight altitude settles at different values from flight to flight, and is more stable during the short time interval of a single flight. Such drone's behavior requires scanning a sufficiently large area to ensure beam inclusion, at the cost of reduced sampling density to maintain a measurement duration compliant with the flight endurance. The color-coded line of Figure 10b indicates the received power after the front-end, in the merged coordinates (UV grid). Samples within compression have been discarded.

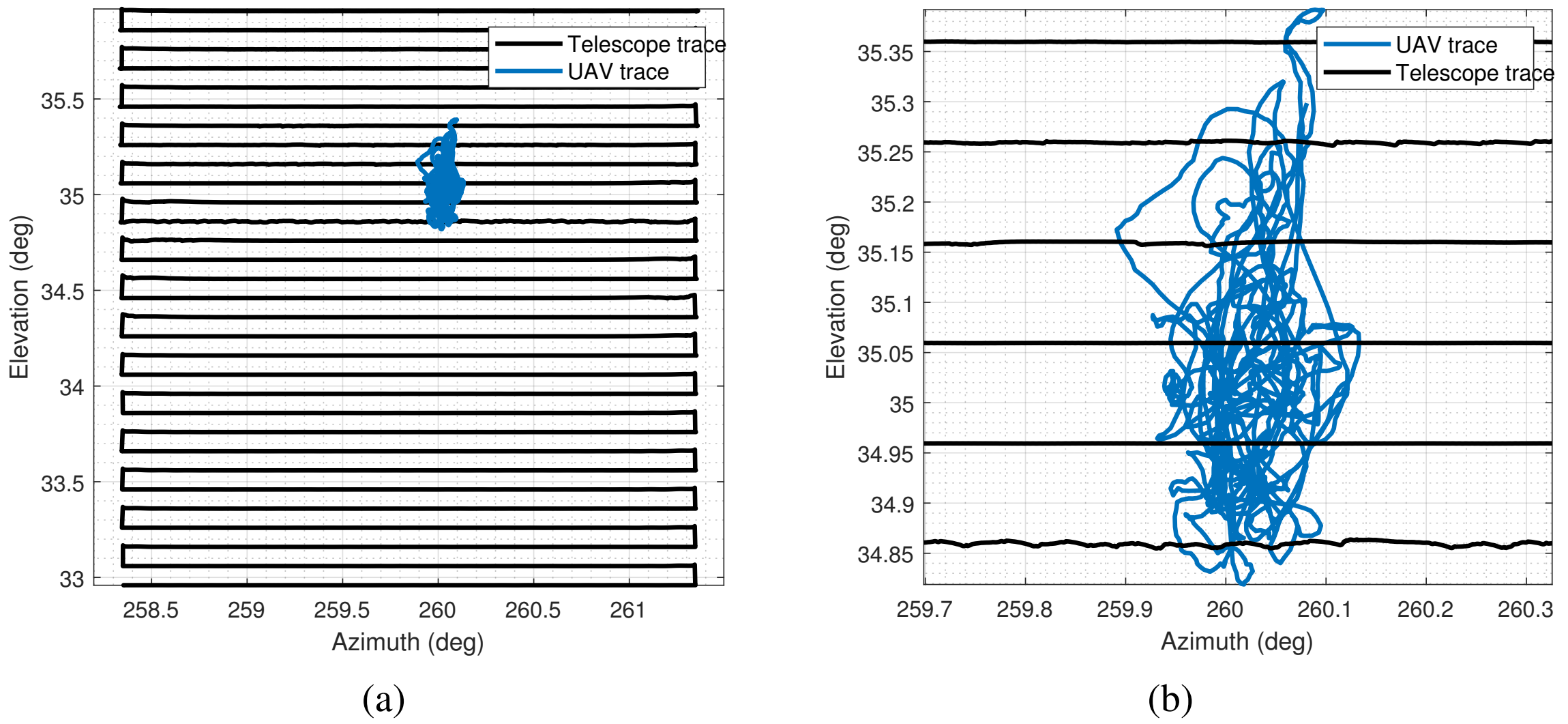

**Figure 7**. Elevation-over-azimuth plot at two zoom levels of telescope scan (black trace) compared with the UAV position throughout the measurement (blue trace).

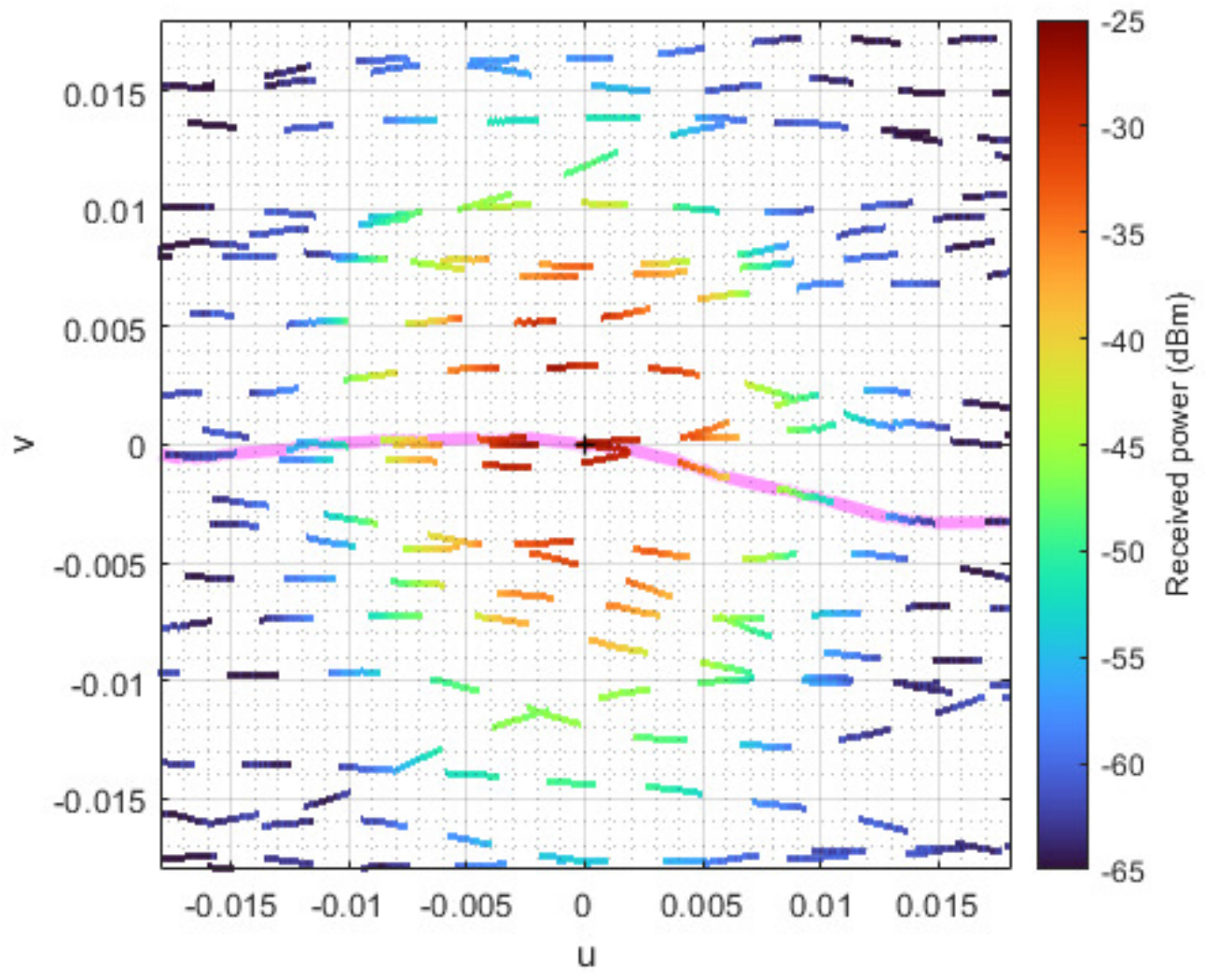

**Figure 8**. Color-coded line in the UV-plane representing the power samples acquired by the spectrum analyzer. The epoch passing closest to the optical axis (the point 0,0 in the UV plane) is highlighted in pink, along with the epochs recorded immediately before and after it.

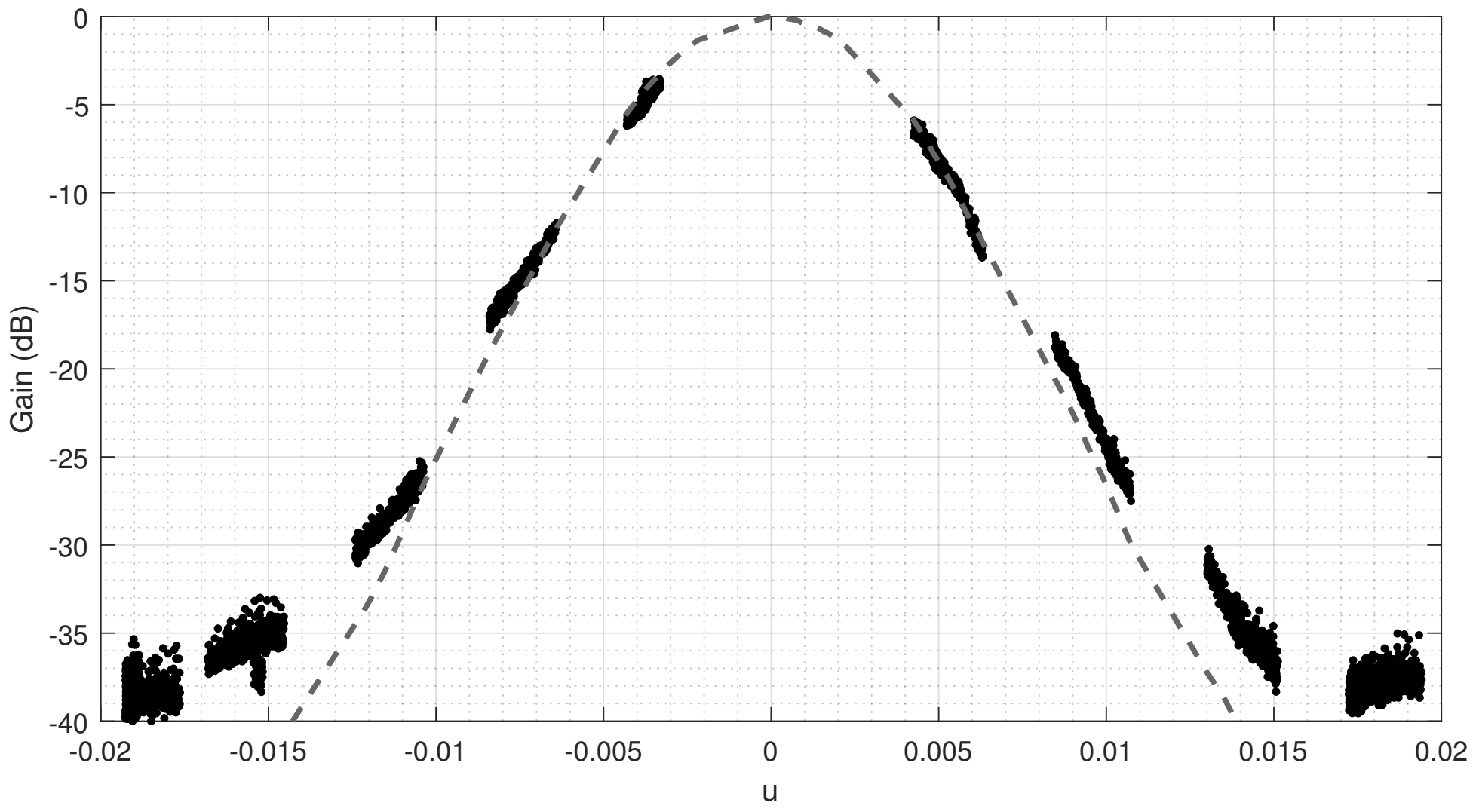

**Figure 9**. Measured (black samples) and simulated (dashed gray trace) gain for feed #1 along the cut highlighted in pink in Figure 8, approximately aligned with the H-plane (hereafter referred to as quasi-H-plane).

The merged dataset is used to compute the colormap of Figure 11a showing the measured beam in a continuous UV plane. Simulation on the same plane is instead shown in Figure 11b. It is important to note that the measured map is generated through linear interpolation of the measured samples into a dense, regular grid. However, the sampling density is higher horizontally and more variable vertically. This feature is somewhat hidden in the maps of Figure 11. However, by merging various acquisitions the measured dynamic range has been increased from 40 dB to 60 dB. A moderate beam ellipticity can be observed, which will be discussed in detail in Section 6.

## 6 Results at Polarimeter Outputs

Using a linearly polarized signal source enables optimal discrimination of the test signal from the background, thereby increasing the Signal-to-Noise Ratio (SNR) and ultimately enhancing the dynamic range of the measurement. When measuring at the output of the polarimeters, the linear components of Stokes parameters, $Q$ and $U$, are computed by properly combining the measured diode-detector voltages, according to Table 1 of [13], from which:

$$Q = \frac{V_{d1} - V_{d2}}{2} \tag{6.1}$$

$$U = \frac{V_{d3} - V_{d4}}{2} \tag{6.2}$$

These formulas assume a linear response and identical behavior across diodes. A more accurate modeling would account for diode response differences and nonlinearity; however, this approach would require individual diode calibration, which is not available. Consequently, for this measurement campaign, the simpler model was adopted, equalizing the diode responses at sky level, i.e. when the telescope is not pointing at the UAV.

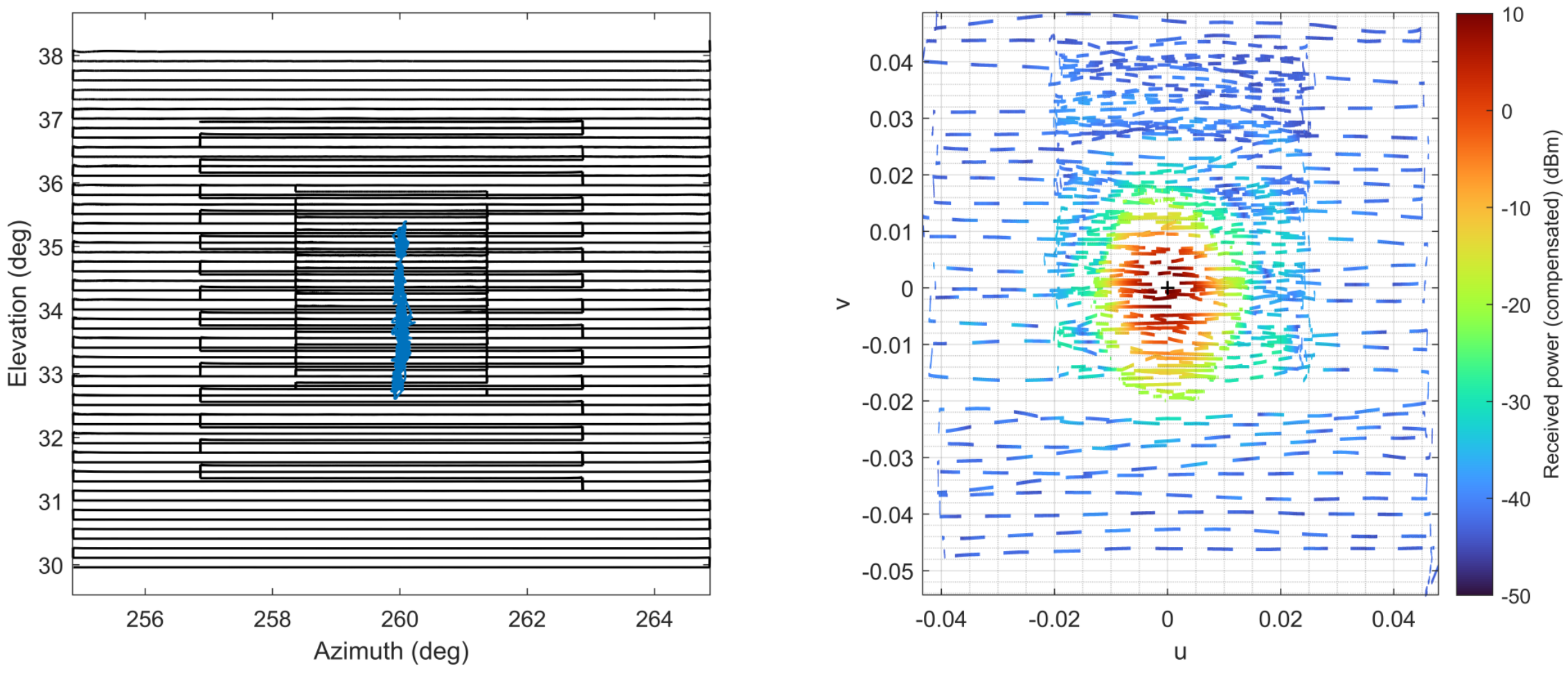

(a) (b)

**Figure 10**. (a) Overall elevation-over-azimuth plot and (b) color-coded line in the UV-plane with data from six different flights overlaid.

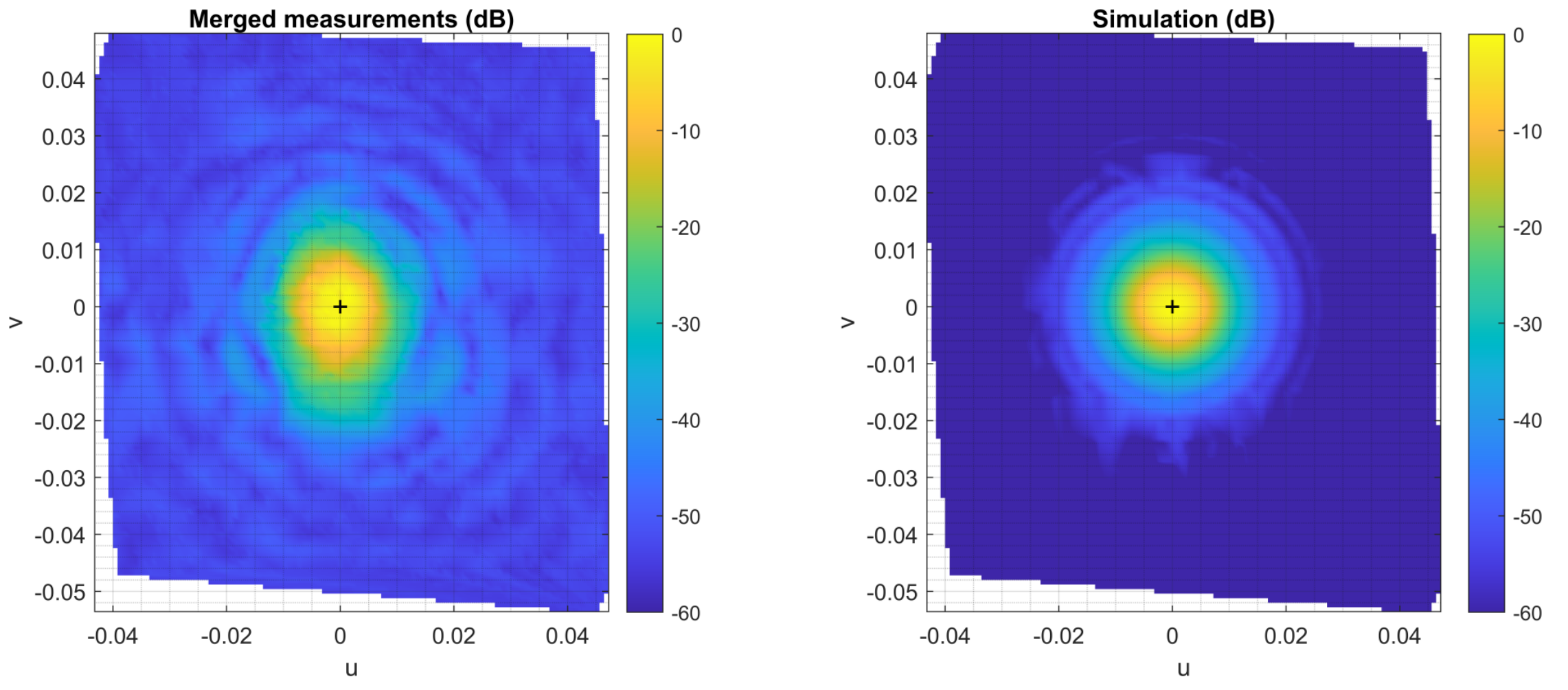

(a) (b)

**Figure 11**. Colormap representation in the UV-plane of measured (a) and simulated (b) gain, linearly interpolated from the set of six flights of Figure 10. Values are normalized.

Figure 12 shows a color-coded line representing the total intensity measured for feed #1 at 33 GHz. As expected, the dynamic range is limited to 6 dB. Noteworthy, lines appear continuous because the data acquired by the telescope consist of an uninterrupted series of samples over time, however, the spacing is variable due to the UAV fluctuation in height. The azimuth rotation speed was 2 degrees per second. The pink highlight indicates the series of samples from which the curves represented in Figure 13 are computed. In particular, the total intensity (black trace) is reported along with the two linear components (solid red and blue lines) and their corresponding simulations (dashed blue and red lines). The level of agreement is already remarkable, and a dynamic range of about 30 dB was achieved. Figure 14 presents instead a direct comparison between the measurements obtained at the front end and with the TFGI, shown in Figure 9 and 13, respectively, demonstrating consistency.

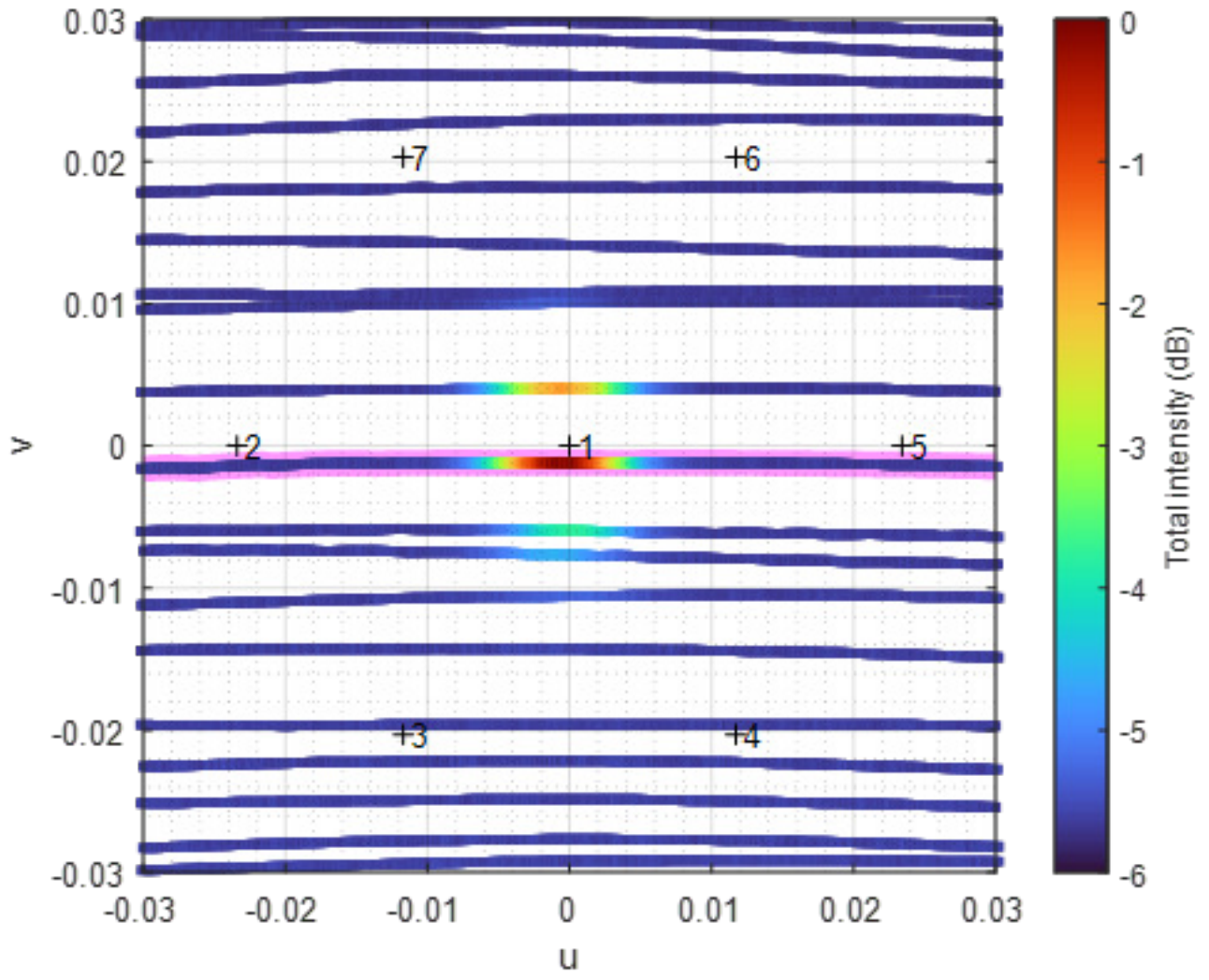

**Figure 12**. Color-coded line in the UV-plane of the linear component Q observed with the TGI, values normalized. Labeled crosses indicating the nominal pointing directions of the feeds.

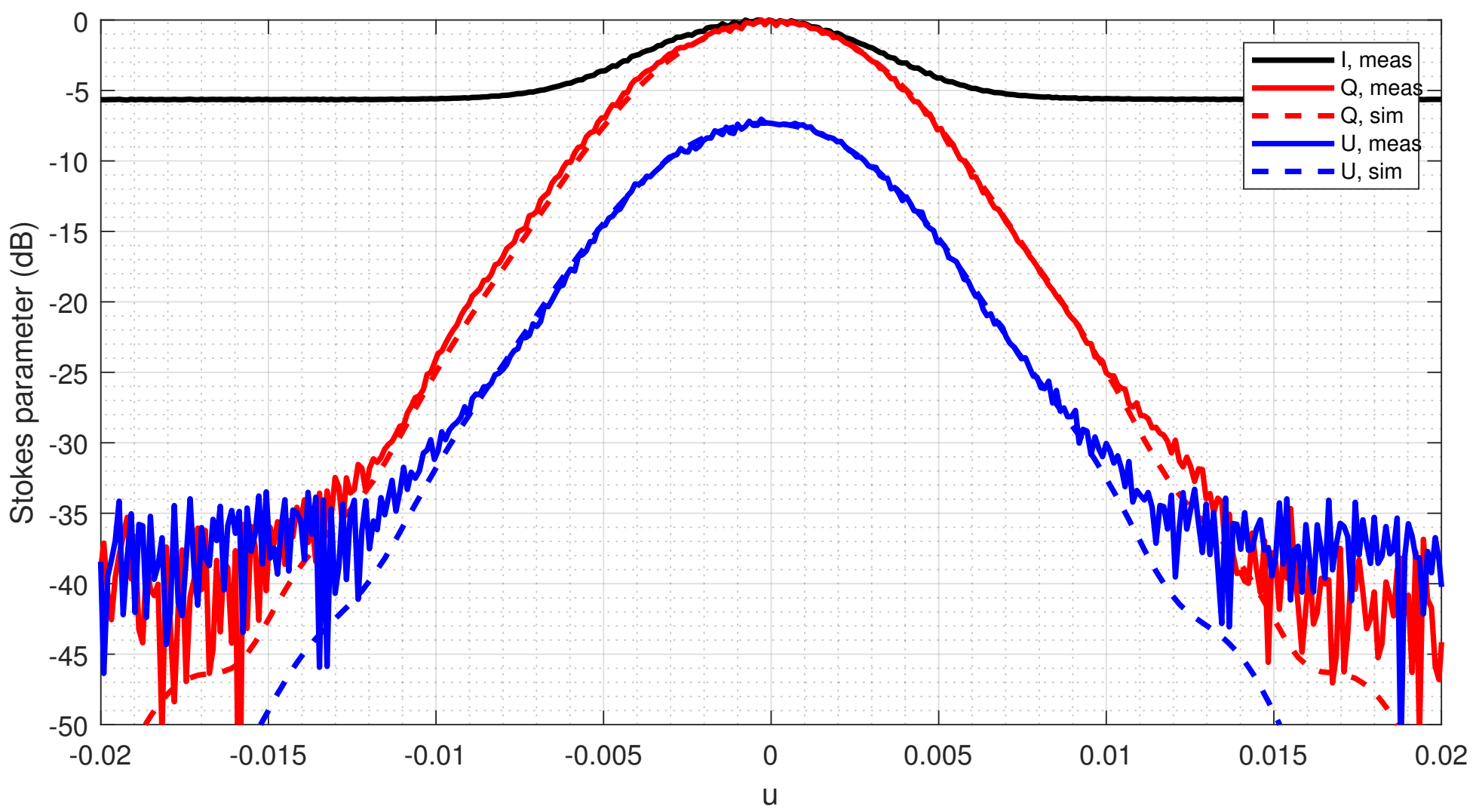

**Figure 13**. Measured (solid lines) and simulated (dashed lines) Stokes parameters for feed #1 along the quasi-H-plane cut highlighted in pink in Figure 12, values normalized. Total intensity I (black), linear components Q (red) and U (blue).

As mentioned in Section 3, the polarization of the radiated signal is linear and nominally lies in the vertical plane. More precisely, the polarization angle coincides with the roll angle of the UAV, $\gamma$, defined with respect to the vertical direction. Given that $\gamma$ is recorded throughout the measurement, the artificial source allows to estimate the rotation angle between the instrumental system and the physical reference system. Specifically, the polarization angle inferred from the measured Stokes $Q$ and $U$ parameters,

$$\psi_{\text{meas}} = \frac{1}{2} \arctan\left(\frac{-U_{\text{meas}}}{Q_{\text{meas}}}\right), \tag{6.3}$$

is compared to the expected one, $\gamma$. Their difference provides an estimate of the instrumental rotation angle,

$$\theta = \psi_{\text{meas}} - \gamma, \tag{6.4}$$

that captures the cumulative effect of residual phase imbalances and geometrical misalignments in the signal path, and it is used to compute the simulated curves consistently. It should be noted that the accuracy of the estimate is on the order of a few degrees, as determined by the standard IMU integrated in the adopted UAV, however, the integration of advanced sensors [35] would allow to develop a dedicated and more accurate platform for this purpose.

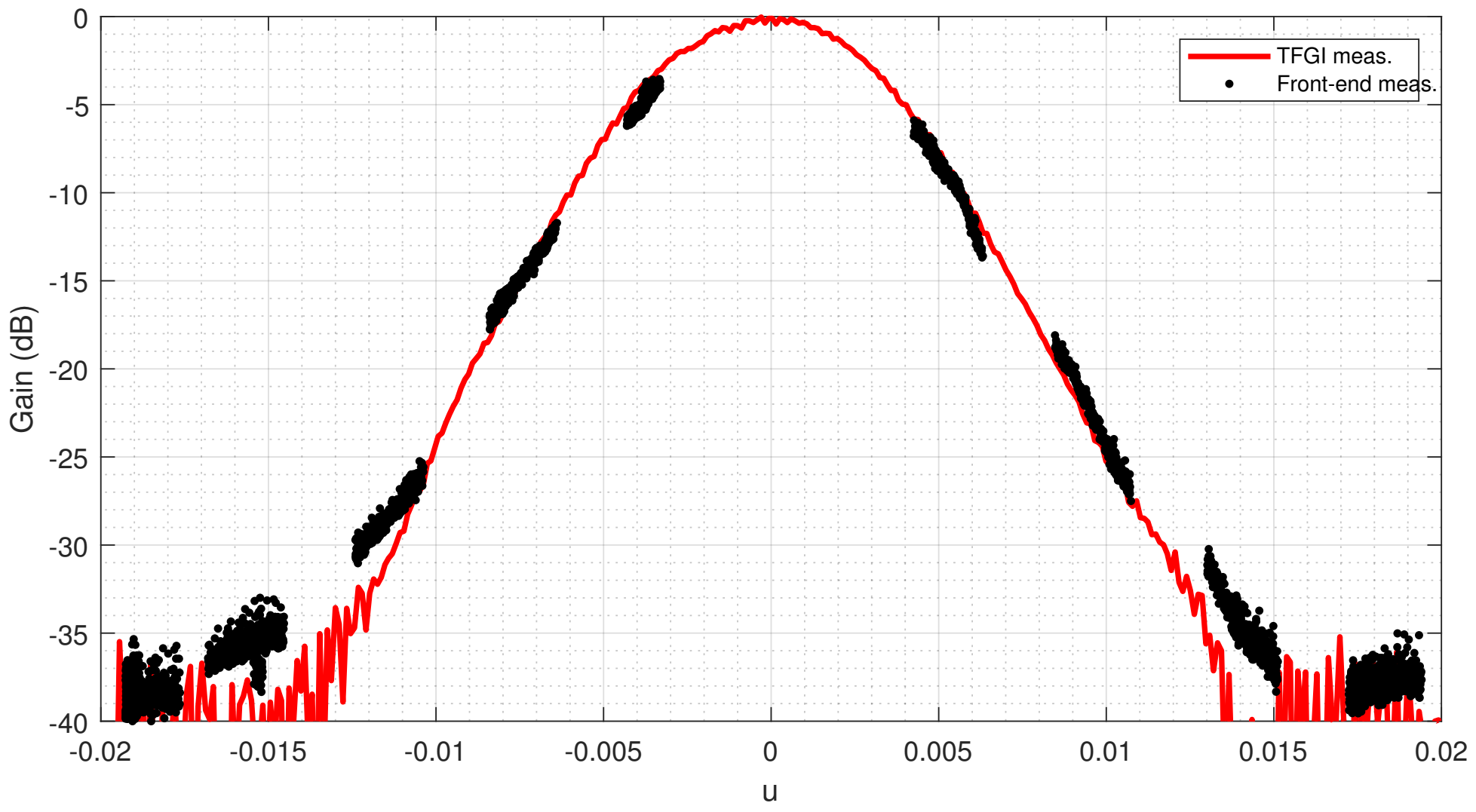

**Figure 14**. Comparison between the quasi-H-plane beam profile measured at the front-end (black markers) and with the TFGI (red curve).

Figure 15 shows instead six colormap representing both measured (upper row) and simulated (lower row) beams for feeds #1, #3 and #7 (first, second and third column, respectively), i.e., all the feeds that were simultaneously receiving at 33 GHz. Also in this case, a moderate beam ellipticity is observed. Figure 16 finally shows an overview of the obtained results for feeds #4, #5 and #6 (first, second and third column, respectively) at 40 GHz.

As noted from the front-end level measurements, a moderate beam ellipticity was also observed at the TFGI output. Figure 17 compares the front-end measurements (cross markers) with those obtained using the TFGI (circle markers). Front-end samples affected by receiver compression were discarded. Error bars indicate the angular uncertainty associated with each acquired sample. This uncertainty reflects the horn's position accuracy in the sky and was conservatively estimated at ±0.028°. This value differs from the uncertainty mentioned in Section 4, i.e., ±0.006°, which accounts only for the horizontal RTK accuracy. The larger uncertainty considered here accounts for the vertical positioning error—assumed to be three times that of the horizontal one, based on experimental evidence [37]—and the UAV attitude uncertainty, since the onboard horn is not located at the drone's center of rotation.
Notably, measurements with the TFGI and at the front-end output are consistent with each other. This cross-comparison allows excluding potential issues introduced by the signal conditioning chain or by the computation of the polarimetric parameters. The cause of the distortion is therefore likely to be at the antenna level or within the measurement system.
The figure also shows the expected beam profile (blue solid line) and a horizontally scaled curve with a factor of ×1.2 (orange solid line). This latter curve serves as a simplified model representing the minimum ellipticity required to fit the measured samples. Based on this, the measured FHPBW is 0.43°, compared to the simulated value of 0.36°, indicating a beam ellipticity with an axial ratio of approximately 1.2.

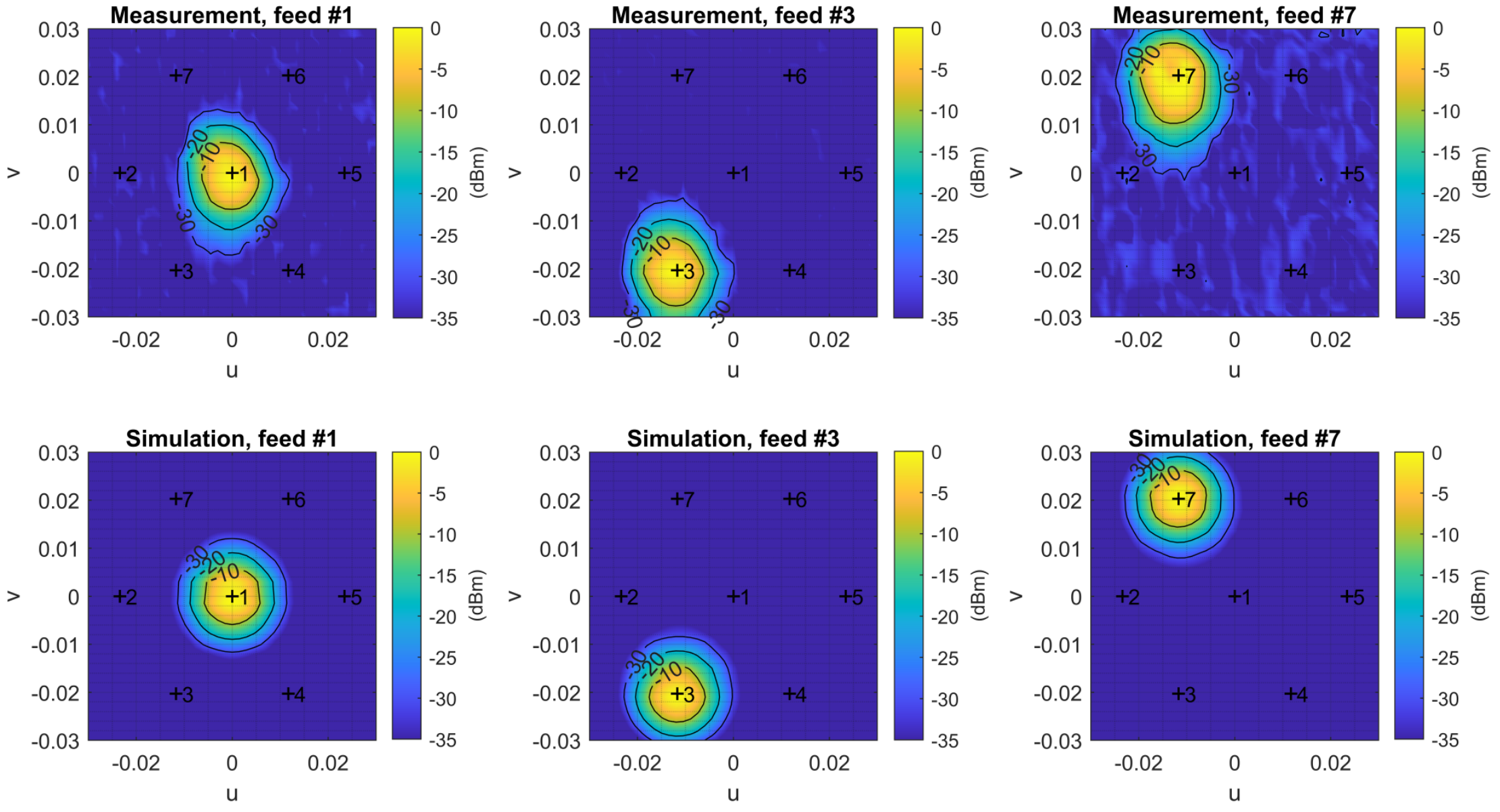

**Figure 15**. Colormap representation in the UV-plane of measured (top row) and simulated (bottom row) linear component Q at 33 GHz for feeds #1 (left column), #3 (center column), and #7 (right column). Cross markers indicating the nominal feed position, contour lines denoting levels –10 dB, –20 dB, and –30 dB.

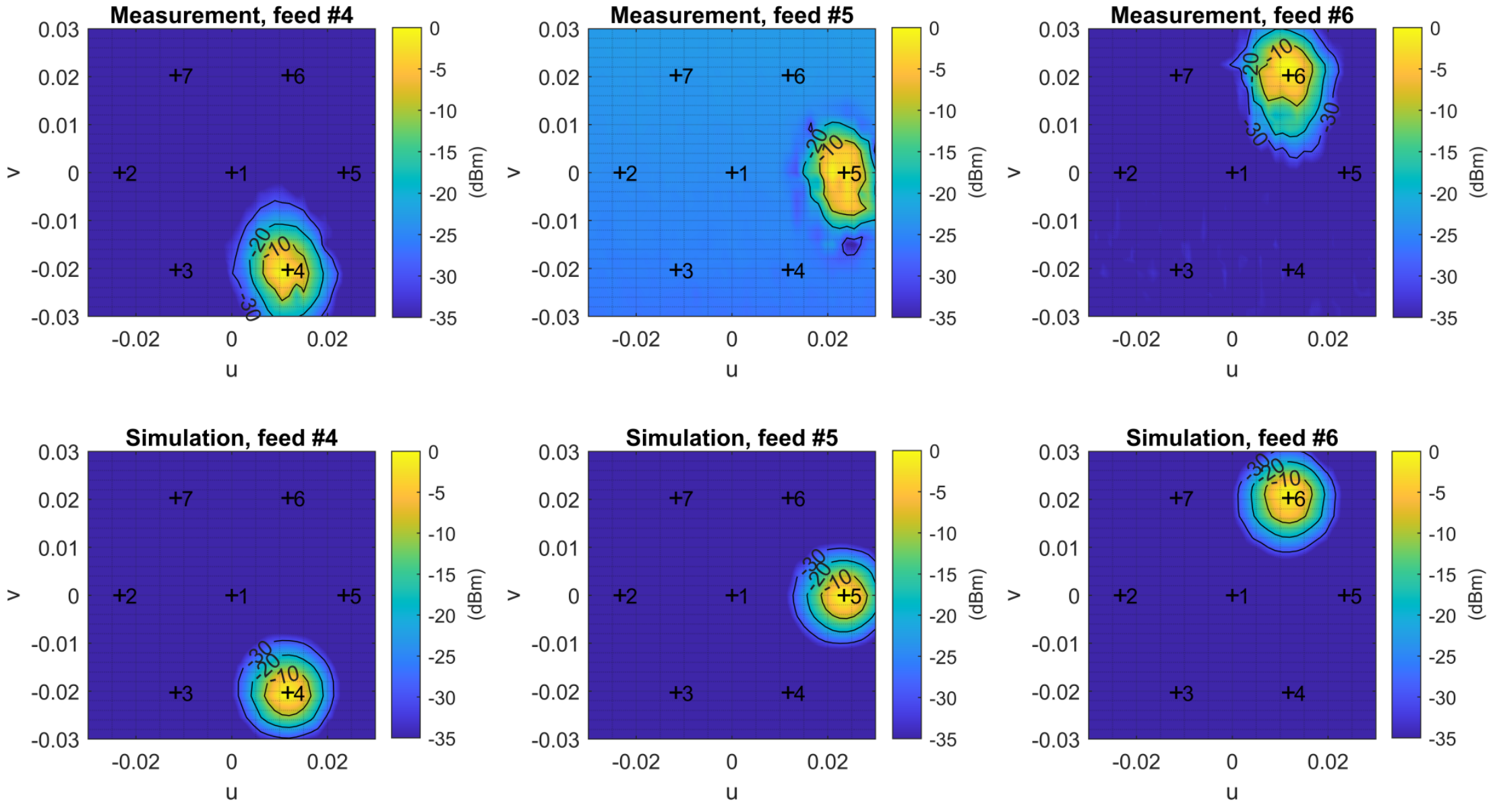

**Figure 16**. Colormap representation in the UV-plane of measured (top row) and simulated (bottom row) linear component Q at 40 GHz for feeds #4 (left column), #5 (center column), and #6 (right column). Cross markers indicating the nominal feed position, contour lines denoting levels –10 dB, –20 dB, and –30 dB.

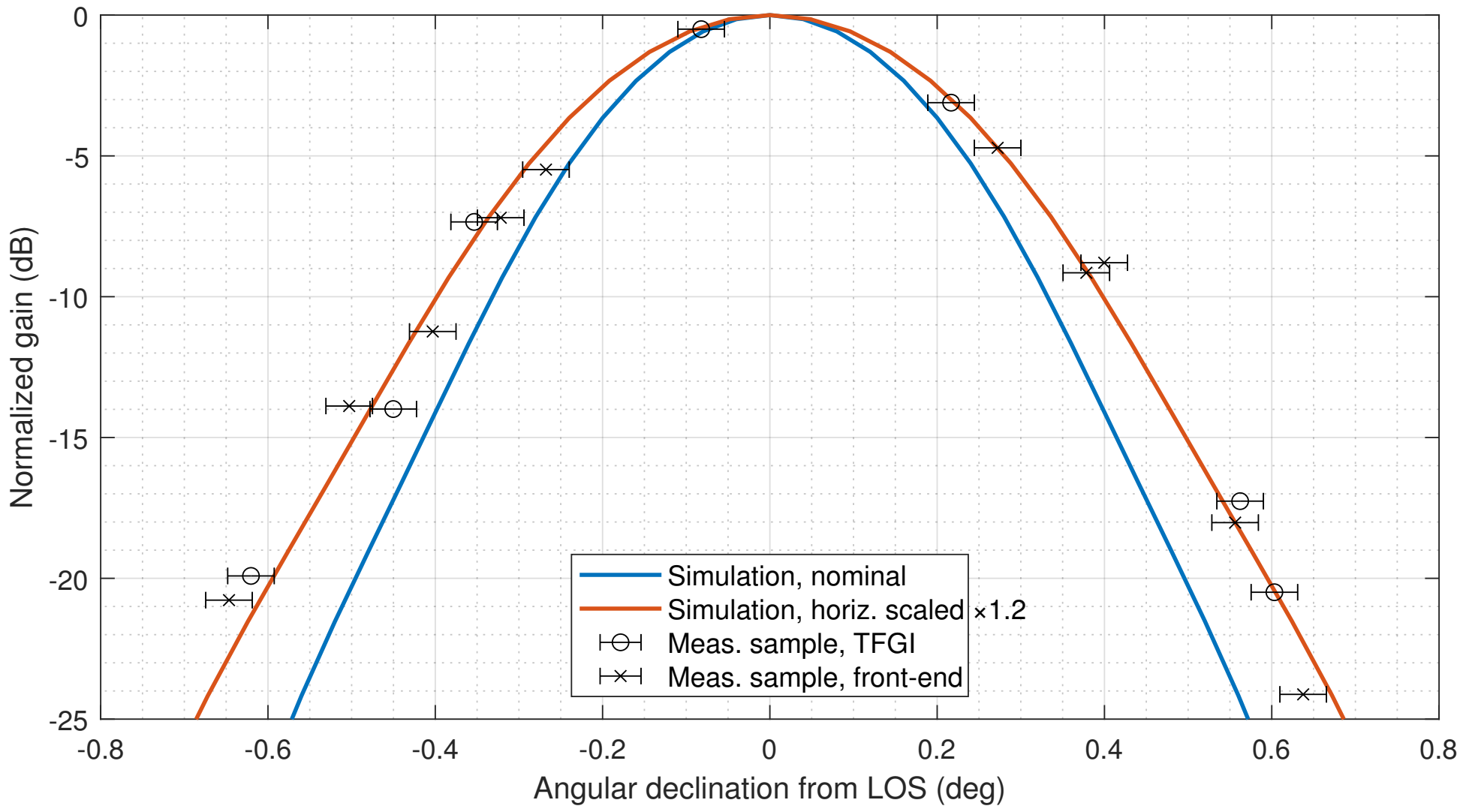

**Figure 17**. Measured samples with TFGI (circle markers) and at the front-end output (cross markers). Error bars represent the angular uncertainty due to UAV pose errors. Nominal simulation (blue curve) and horizontally scaled simulation ×1.2 (orange solid line) are also shown.

The observed elongation develops in the vertical direction. Since the UAV during flight also exhibits a more pronounced fluctuation in the vertical axis, compared to the horizontal plane, we considered the possibility that the observed effect in the measurement could be attributable to the drone itself. However, we emphasize that RTK measurements are available for all three spatial coordinates, although the UAV altitude navigation is performed with the barometer. Prior work employing a less precise sensor (namely a u-blox M8P RTK having a nominal accuracy of 2.5 cm [38]) has corroborated the expected level of accuracy through cross-validation with near-field phase measurements [39]. These measurements rely on spatially accurate sampling to extrapolate the far-field radiation pattern from measurements taken in the near-field region. With reference to Figure 17, the positioning uncertainty required to intercept the nominal simulation is on the order of 30 cm, which is not consistent with the aforementioned experiments and other evidences [37].

Furthermore, previous validation tests of the drone platform, involving the measurement of a reference horn's E-plane radiation pattern (i.e. the same cut where ellipticity has been observed during the campaign on QUIJOTE), did not reveal any comparable distortion [8]. An additional consideration pertains to the potential overestimation of height fluctuations by the differential GPS. However, this hypothesis is not corroborated by the data trace obtained during the determination of the telescope's center of rotation (see details in Appendix A), which on the contrary exhibits a remarkable stability.

Above all, the presence of UAV-related position errors is expected to cause irregular beam distortion and/or displacement of the beams from their nominal positions. However, the beam centering observed in Figure 15 and 16 matches the simulations.

We also considered potential inaccuracies in the formulation of the transformation between reference systems (see Appendix B). However, when strictly focusing on the vertical plane containing the telescope's line-of-sight (LOS), the reference system transformation formulas play no role. In other words, within the vertical plane, the scan inherently occurs along a principal axis (the V-axis), and the transformation formulas do not introduce any scaling correction.

A further consideration concerns the accuracy of the altitude estimate for the telescope's rotation center (see Appendix A). However, such an error would have led to a clearly incorrect elevation of the drone in the sky, and consequently to a significant apparent elevation pointing error of the beams, not to an elongation.

A final consideration pertains to possible RFI issue from the drone. Most emissions typically originate from unshielded onboard electronics—such as power supplies, motor electronic speed controllers, and similar components—which predominantly emit at much lower frequencies (below 1 GHz). Furthermore, since the used scanning strategy involves the drone remaining stationary, any drone-generated RFI would not account for a distortion occurring only in the vertical plane.

Although the body of evidence discussed above reasonably supports the exclusion of systematic effects intrinsic to the measurement system (the UAV) as the cause of the observed phenomenon, it is nevertheless notable that observations of celestial sources with QUIJOTE-TFGI have not previously exhibited an ellipticity as pronounced as that currently observed (axial ratio of 1.2). Indeed, the measured average ellipticities for the central pixel with Tau A observations are on the order of 1.03 of axial ratio at –3 dB. This discrepancy requires further investigation. Future tests, as outlined in the Section 7, will address this issue and aim to provide a comprehensive understanding of the observed beam characteristics.

## 7 Campaign Outcomes and Forecasts

The primary objective of the experimental campaign on QUIJOTE-TFGI was the validation of both modeling and testing approaches, i.e., using a drone-based source to measure the beams and verify them against near-field models. Preliminary data from this campaign achieved remarkable accuracy, but also revealed unexpected results that require further investigation through additional experimental data collection. The particular feature of QUIJOTE, which allows collecting data both from the polarimeters and at the front-end output, provided an excellent opportunity for double-checking the beam using two different receivers on the same antenna.

A preliminary measurement campaign, such as the one conducted at QUIJOTE, requires substantial preparatory activities prior to effective data acquisition. Future campaigns will benefit from the implementation of a wider range of scanning strategies. For instance, the telescope could employ vertical raster scans (elevation scans with azimuthal stepping) to achieve finer beam sampling also in the vertical direction. Furthermore, the drone platform could be enhanced with advanced sensors, such as an RTK altimeter, mitigating the intrinsic limitations of barometric altitude navigation. This upgrade would enable a significant reduction in the scan area, tailored to the beam, thus facilitating denser sampling. Alternatively, performing drone scans while the telescope remains stationary would provide additional data for the characterization of systematic effects, including static and dynamic pointing errors. Features observed during this campaign, such as beam ellipticity, could be investigated more thoroughly and fully understood through comparative analysis of multiple measurements.

Moreover, the integration of RTK-aided inertial sensors with sub-tenth-degree accuracy [35] would enable the calibration of the Q/U basis with two order of magnitude improvement in accuracy compared to the current campaign. Recent studies have proposed integrated systems employing photogrammetry [40, 41], a technique enabling dimensional and positional measurements of objects through image acquisition and analysis, achieving accuracies on the order of 0.1 degree. RTK-aided IMUs further enhance accuracy while eliminating the need for photogrammetric data analysis.

The forthcoming construction of LSPE-Strip near QUIJOTE will enable joint observations, allowing for cross-verification between the two instruments, enhancing the reliability and robustness of the data. LSPE-Strip has a wider focal plane with 49 correlation polarimeters in Q-band and 6 in W-band, and the current configuration of QUIJOTE-TFGI has 19 pixels. This is an additional challenge in terms of scan strategy with respect to QUIJOTE where 7 pixels (30 GHz and 40 GHz) were available in 2022.

Finally, it is important to mention that the LSPE-Strip pointing model will be calibrated using an optical camera (i.e., the Star Tracker). This device will refine the model's accuracy by determining the true pointing direction based on motor encoder positions. Due to the unavoidable misalignment between the optical axes of the Star Tracker and LSPE-Strip, a specialized calibration procedure is required. This procedure will run in parallel with the antenna beam pattern characterization, correlating the drone's position (captured by the Star Tracker image) with the measured beam pattern. Since the drone will occupy only a few pixels in the Star Tracker image, it will be equipped with a light source, providing a precise reference point for accurate position identification [9].

## A Determination of the Telescope's rotation center

The RTK system requires the presence of a ground receiver, known as the base station, and a second receiver on the UAV, referred to as the rover. The base station is connected to the Ground Control Station (GCS), which consists of a computer running the flight management software that facilitates real-time monitoring and control of the UAV mission. Both the base station and the GCS are typically situated near the home location (i.e., takeoff and landing points). The base station assumes its geographic position (i.e., latitude, longitude, and altitude) to be known and fixed. Through telemetry link enabling bidirectional communication between the GCS and the UAV, the base station streams both GNSS observations and its assumed position to the rover. The rover then combines its own observations with those from the base station to compute the RTK solution. This setup allows for high positional accuracy of the UAV in the local reference system (relative to the base station). In other words, every differential GPS system builds a local reference system centered on the base. Interestingly, the assumed geographic position of the base can be arbitrarily chosen. The local coordinate system is Cartesian, with the x-axis directed eastward (tangent to the parallel), the y-axis directed northward (tangent to the meridian), and the z-axis pointing upward (commonly referred to as the East-North-Up, ENU system).

For the purpose of antenna measurement, it is essential to determine the UAV coordinates relative to the Antenna Under Test (AUT). This passage requires knowledge of the AUT position in the local reference system and, of course, obtaining this datum is crucial. Advanced tools such as laser trackers can be employed with some difficulty in outdoor environments. Further complications represented by the distance between the AUT and the base station, presence of obstacles, and solar radiation. Another method consists in acquiring, with the RTK system, a particular point from which the AUT location can be easily derived. For this measurement campaign, the following procedure was employed with a small change: the UAV was positioned above a rack cabinet fixed to the telescope alongside the aperture (see Figure 18). The telescope was then set to an elevation of 90° (i.e., pointing toward the zenith) and continuously rotated in azimuth by 360° in 8 minutes. During the rotation, the RTK trace measured by the UAV was recorded. The blue circular trace illustrated in Figure 19a allowed for the determination of the vertical rotation axis position, which coincides with the center of the circle that best fits the acquired trace (black dashed). The eccentricity (Figure 19b) has a root-mean-square (RMS) value of about 5 mm consistently with the RTK nominal accuracy. The height of the telescope's elevation axis was instead derived based on the measured mean height (Figure 19c) and technical drawings of the instrument.

## B Reference Systems and Transformations

The measurement of mechanically steered telescopes using drones involves the simultaneous motion of two objects relative to each other. This necessitates performing sample-wise transformations between the different reference systems employed by each tool. The following procedure outlines the operations performed:

- UAV position: the drone's coordinates are computed in the East-North-Up (ENU) Cartesian reference system, relative to the telescope's rotation center (see Appendix A). These coordinates represent the center of the drone's frame and not the horn mounted on the UAV.

- True horn position: UAV position data are processed to account for the UAV's attitude angles (yaw, pitch, and roll) and the horn displacement within the UAV frame. The output is the horn position expressed in the ENU reference system.

- Rotated source position: telescope's azimuth and elevation angles are applied sample-wise to the true source position $(x, y, z)$ to derive the UAV position in the telescope-fixed system, $(x', y', z')$. These rotations are described by the rotation matrices (B.1) where $\alpha$ is the telescope's declination, $\gamma$ is the azimuth, and $R_y = I$ (identity matrix) since the telescope does not rotate about the y-axis. Rotations are then applied with (B.2) (elevation after azimuth).

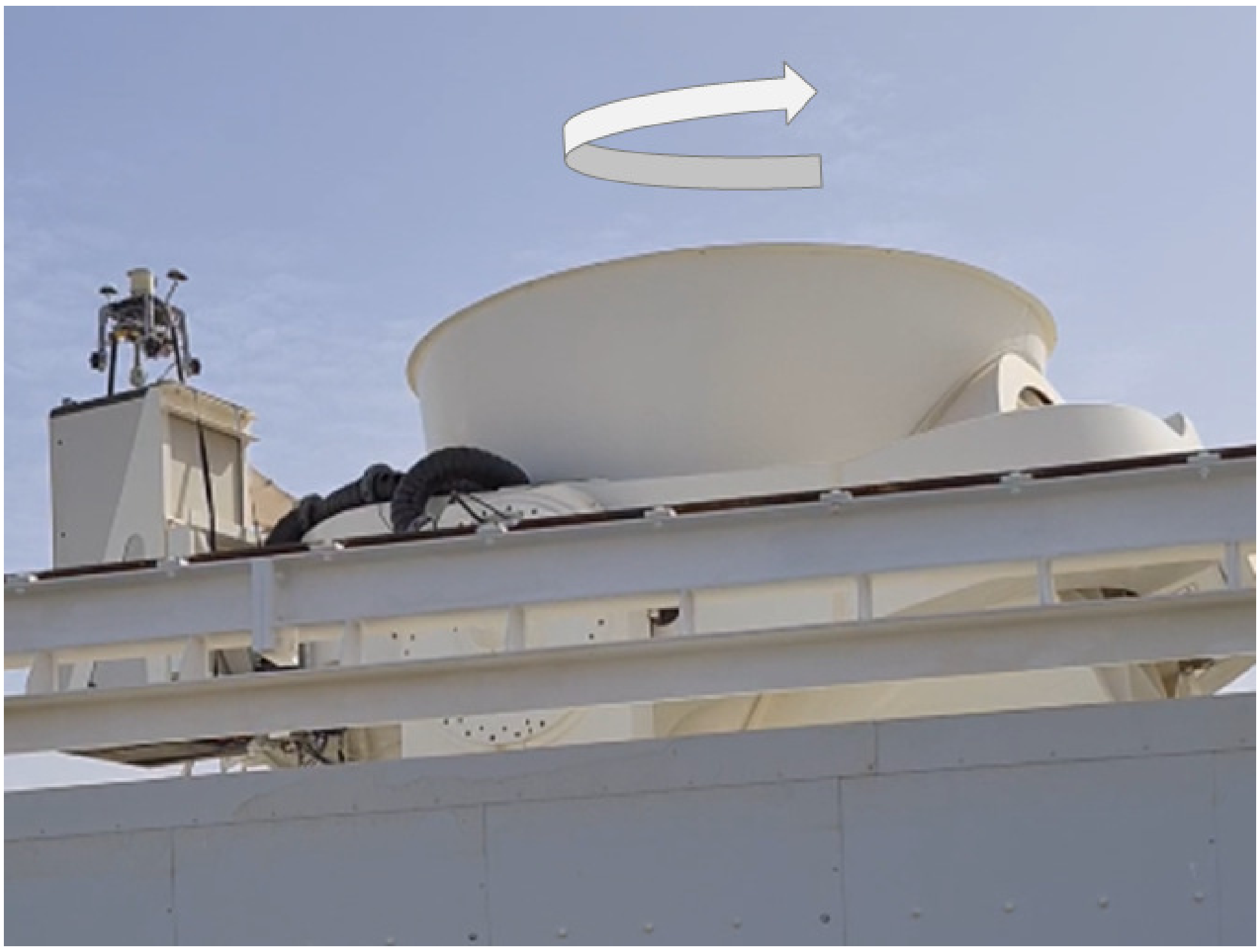

**Figure 18**. The UAV fastened on a rack cabinet alongside the telescope to acquire the azimuth rotation center.

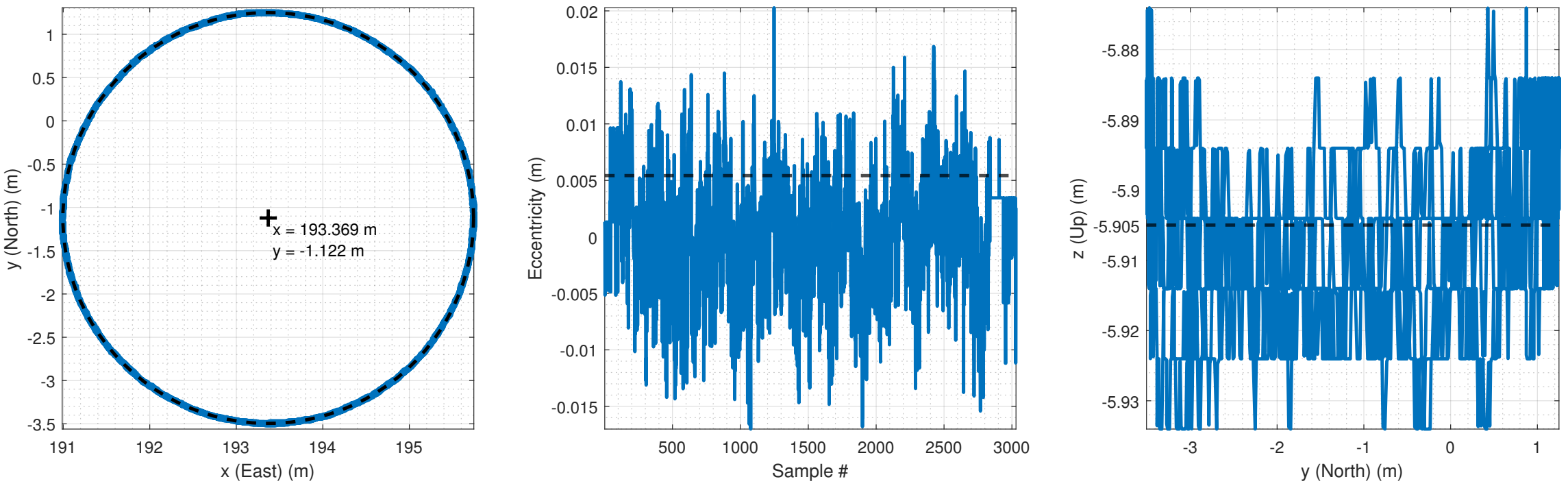

(a) (b) (c)

**Figure 19**. Recorded RTK traces (blue lines) during telescope complete rotation. The curves enable to determine the 3D position of the telescope's center of rotation relative to the UAV's home location. (a) Projection on the horizontal plane, dashed line and cross indicating best fitting circle and its center. (b) Eccentricity, dashed line marking the RMS. (c) Projection on the vertical plane, dashed line denoting mean value.

- UV Components: the UV components are derived with (B.3), where $\theta$ and $\phi$ are the angular coordinates of a spherical system; i.e., $\theta$ is the angular displacement of the UAV from the LOS (polar angle), and $\phi = \arctan(y'/x')$.

$$R_x(\alpha) = \begin{bmatrix} 1 & 0 & 0 \\ 0 & \cos\alpha & -\sin\alpha \\ 0 & \sin\alpha & \cos\alpha \end{bmatrix},$$

$$R_y(\beta) = \begin{bmatrix} \cos\beta & 0 & \sin\beta \\ 0 & 1 & 0 \\ -\sin\beta & 0 & \cos\beta \end{bmatrix} = I, \tag{B.1}$$

$$R_z(\gamma) = \begin{bmatrix} \cos\gamma & -\sin\gamma & 0 \\ \sin\gamma & \cos\gamma & 0 \\ 0 & 0 & 1 \end{bmatrix}.$$

$$\begin{bmatrix} x' \\ y' \\ z' \end{bmatrix} = R_x(\alpha) R_z(\gamma) \begin{bmatrix} x \\ y \\ z \end{bmatrix} \tag{B.2}$$

$$\begin{aligned} u &= \sin(\theta)\cos(\phi) \\ v &= \sin(\theta)\sin(\phi) \end{aligned} \tag{B.3}$$

A sketch of the telescope-fixed UV grid is shown in Figure 20, where $u$ has the same orientation as the $x'$-axis, $v$ has the same orientation as the $y'$-axis, and the LOS coincides with the $z'$-axis. It is worth noting that $\phi = 0°$ is an equatorial plane whereas the azimuth rotation plane is horizontal. Therefore, assuming the UAV remains stationary, when the telescope rotates azimuthally, the drone will not still lie in the plane $\phi = 0°$. Conversely, both $\phi = 90°$ and the elevation rotation plane are meridian planes; therefore, when the telescope change its elevation the UAV always lies in the plane $\phi = 90°$.

## Acknowledgments

The authors thank Andrea Berton, head of the CNR group responsible for flight operations management.
The *LSPE-Strip* Project is carried out thanks to the support of ASI (contract I/022/11/1 and Agreement 2018-21HH.0) and INFN.
The Strip instrument will be installed at the Observatorio del Teide, Tenerife, of the Instituto de Astrofísica de Canarias.
The Astronomical Observatory of the Autonomous Region of the Aosta Valley (OAVdA) is managed by the Fondazione Clément Fillietroz-ONLUS, which is supported by the Regional Government of the Aosta Valley, the Town Municipality of Nus and the "Unité des Communes valdôtaines Mont-Émilius". Stefano Sartor acknowledges funds from a 2020 'Research and Education' grant from Fondazione CRT-Cassa di Risparmio di Torino.
The *QUIJOTE* experiment is being developed by the Instituto de Astrofisica de Canarias (IAC),

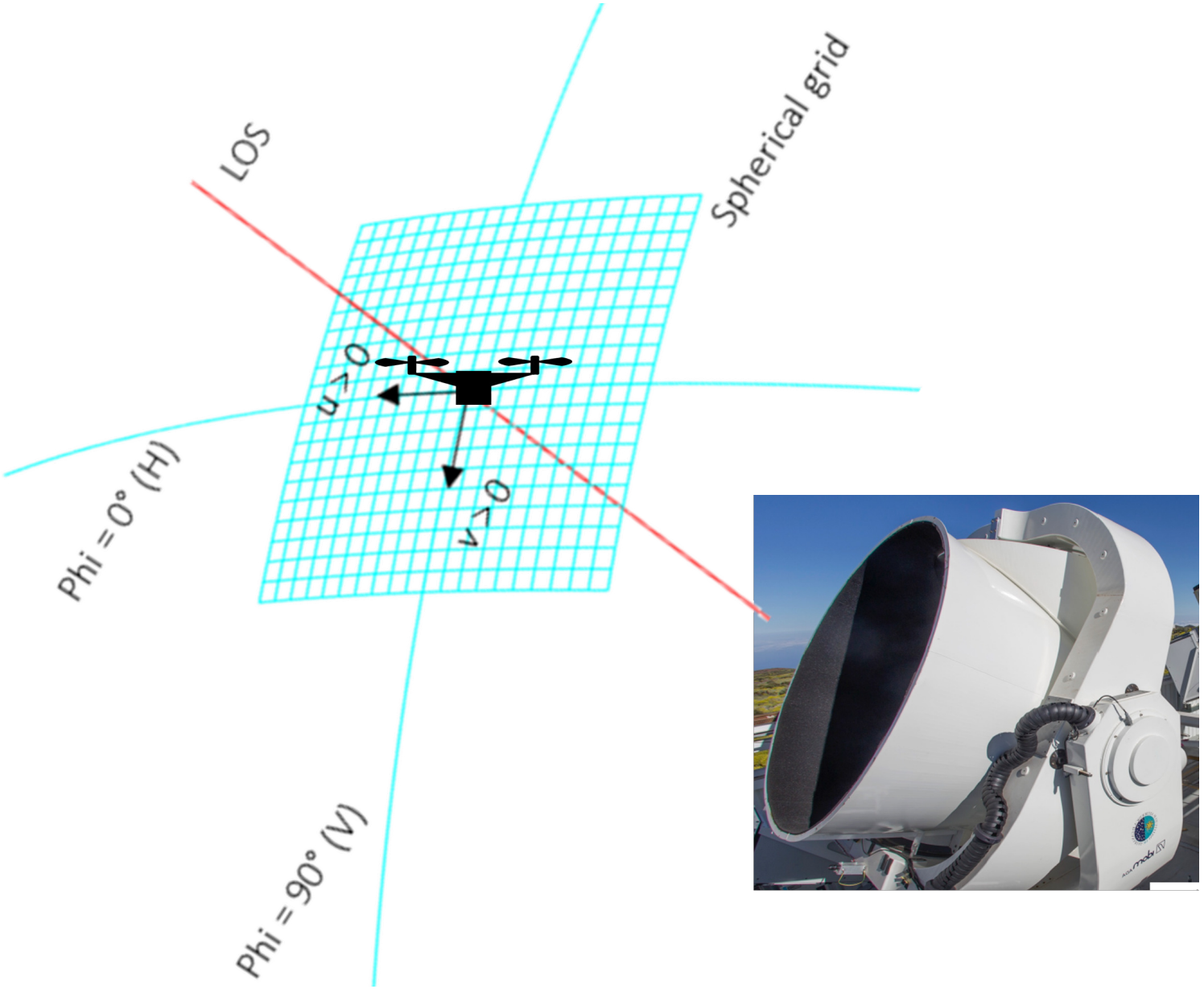

**Figure 20**. Illustration of the reference system. Red line is the LOS of QUIJOTE, coincident with that of feed #1.

the Instituto de Fisica de Cantabria (IFCA), and the Universities of Cantabria, Manchester and Cambridge.
Partial financial support was provided by the Spanish Ministry of Science and Innovation under the projects AYA2007-68058-C03-01, AYA2007-68058-C03-02, AYA2010-21766-C03-01, AYA2010-21766-C03-02, AYA2014-60438-P, ESP2015-70646-C2-1-R, AYA2017-84185-P, ESP2017-83921-C2-1-R, PGC2018-101814-B-I00, PID2019-110610RB-C21, PID2020 -120514GB-I00, IACA13-3E-2336, IACA15-BE-3707, EQC2018-004918-P, the Severo Ochoa Programs SEV-2015-0548 and CEX2019-000920-S, the Maria de Maeztu Program MDM-2017-0765, and by the Consolider-Ingenio project CSD2010-00064 (EPI: Exploring the Physics of Inflation). We acknowledge support from the ACIISI, Consejeria de Economia, Conocimiento y Empleo del Gobierno de Canarias and the European Regional Development Fund (ERDF) under grant with reference ProID2020010108, and Red de Investigación RED2022-134715-T funded by MCIN/AEI/10.13039/501100011033.
This project has received funding from the European Union's Horizon 2020 research and innovation program under grant agreement number 687312 (RADIOFOREGROUNDS), and the Horizon Europe research and innovation program under GA 101135036 (RadioForegroundsPlus).